\documentclass[twocolumn]{autart}    
\usepackage{amsmath,amssymb,amsfonts}
\usepackage{dsfont}
\usepackage{graphicx}
\usepackage{algorithm,algorithmic}
\usepackage{color}
\usepackage{cite}
\usepackage{subcaption}
\usepackage{epstopdf}
\usepackage{bm}
\usepackage{mleftright}
\usepackage{mathrsfs}
\usepackage{stmaryrd}
\usepackage{apacite}

\newtheorem{theorem}{Theorem}
\newtheorem{proposition}{Proposition}

\newtheorem{lemma}{Lemma}

\newtheorem{definition}{Definition}

\newtheorem{remark}{Remark}
\newcommand{\mathbd}[1]{\mbox{\boldmath $#1$}}

\DeclareMathOperator*{\argmin}{arg\,min}
\DeclareMathOperator*{\argmax}{arg\,max}

\def\bmP{\mbox{$\bm{P}$}}
\def\bmC{\mbox{$\bm{C}$}}
\def\bmG{\mbox{$\bm{\Gamma}$}}
\def\bmF{\mbox{$\bm{F}$}}
\def\bmD{\mbox{$\bm{\Delta}$}}
\def\bmT{\mbox{$\bm{T}$}}
\def\bmI{\mbox{$\bm{I}$}}
\def\nlcls{\mbox{$\bm{P}\,\#\,\bm{C}$}}
\def\lcls{\mbox{$P\,\#\,C$}}
\def\Re{\mbox{\rm Re}}
\def\Im{\mbox{\rm Im}}
\begin{document}

\begin{frontmatter}

\title{
Stabilization of Cascaded Two-Port Networked Systems with Simultaneous Nonlinear Uncertainties\thanksref{footnoteinfo}}

\thanks[footnoteinfo]{This work was supported in part by the Research Grants Council of Hong Kong Special Administrative Region, China, under the project 16201115 and Theme-Based Research Scheme T23-701/14N. }

\author[HKUST]{Di Zhao}\ead{dzhaoaa@connect.ust.hk},    
\author[HKU]{Sei Zhen Khong}\ead{szkhongwork@gmail.com},     
\author[HKUST,ost]{Li Qiu}\ead{eeqiu@ust.hk}  
\thanks[ost]{Tel.:~+852-2358-7067 \ \ \ Fax:~+852-2358-1485.}
\address[HKUST]{Department of Electronic and Computer Engineering, The Hong Kong University of Science and Technology, Clear Water Bay, Kowloon, Hong Kong, China}  
\address[HKU]{Independent researcher}


\begin{abstract}                          
We introduce a versatile framework to model and study networked control systems (NCSs). An NCS is described as a feedback interconnection of a plant and a controller communicating through a bidirectional channel modelled by cascaded nonlinear two-port networks. This model is sufficiently rich to capture various properties of a real-world communication channel, such as distortion, interference, and nonlinearity. Uncertainties in the plant, controller and communication channels can be handled simultaneously in the framework. We provide a necessary and sufficient condition for the robust finite-gain stability of an NCS when the model uncertainties in the plant and controller are measured by the gap metric and those in the nonlinear communication channels are measured by operator norms of the uncertain elements. This condition is given by an inequality involving ``arcsine'' of the uncertainty bounds and is derived from novel geometric insights underlying the robustness of a standard closed-loop system in the presence of conelike nonlinear perturbations on the system graphs.

\end{abstract}

\end{frontmatter}

\section{Introduction}
Feedback is widely used for handling uncertainties in the area of systems and control. Within a feedback loop, communication between the
plant and controller plays an important role in that the achieved control performance and robustness heavily rely on the quality of communication. In
practice, communication can never be ideal due to the presence of channel distortion and interference. Most control systems can be regarded as structured networks with signals \textcolor{black}{transmitted} through channels powered by various devices, such as sensors or
satellites. This gives rise to networked control systems (NCSs), which differ from standard closed-loop systems in that the information is exchanged through communication
networks \cite{zhang2001stability}. The presence of non-ideal communication may introduce disturbances to a control system and hence significantly compromise its
performance.

\begin{figure}
  \centering
  \includegraphics[scale=0.58]{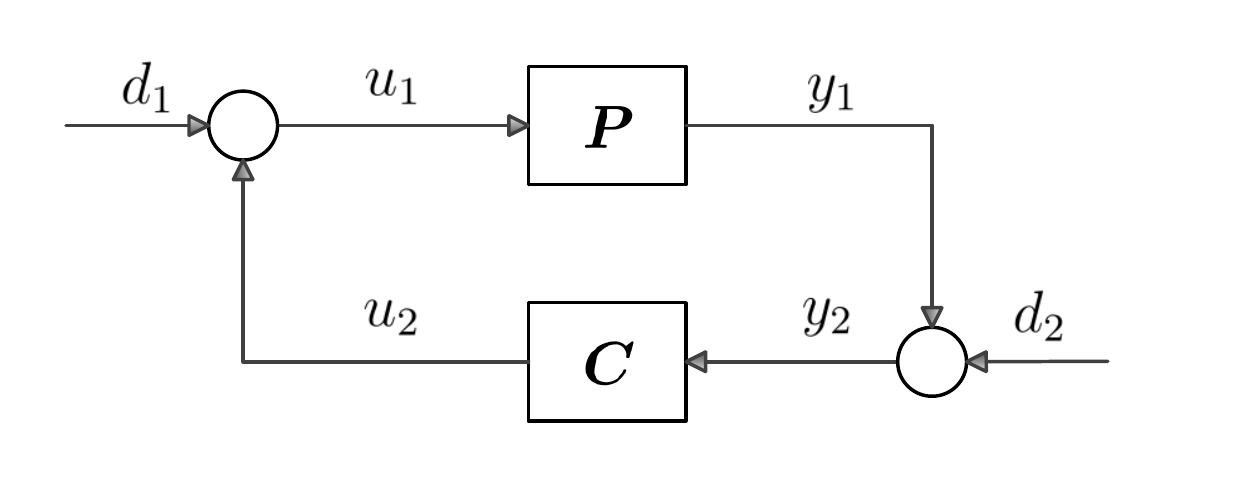}\\
  \caption{A standard closed-loop system.}\label{figloop}
\end{figure}

\begin{figure}
  \centering
  \includegraphics[scale=0.58]{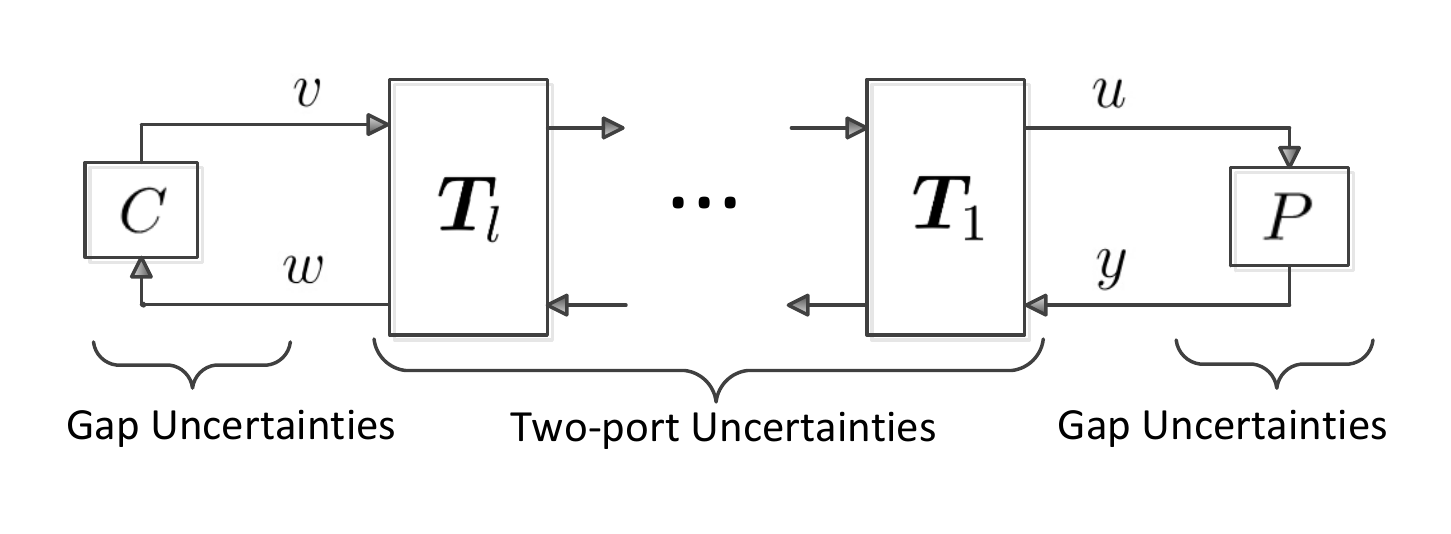}\\
  \caption{Two-port NCS with uncertainties.}\label{fignetwork}
\end{figure}

In this study, we introduce a two-port NCS model, \textcolor{black}{generalizing} a standard finite dimensional linear time-invariant (FDLTI) closed-loop system (Fig.~\ref{figloop}) to \textcolor{black}{a} feedback
system with cascaded two-port connections (Fig.~\ref{fignetwork}). Therein, the plant $P$ and controller $C$ are uncertain FDLTI systems that may be open-loop unstable while the perturbations on the transmission matrices $\bmT_k$ of the two-port communication networks are nonlinear. It is known that model uncertainties are well characterized through the gap metric and its \textcolor{black}{variants}, among which the gap \cite{zames1980proc,georgiou1988computation,georgiou1990optimal,qiu1992feedback}, the pointwise gap \cite{schumacher1992pointwise,qiu1992pointwise} and the $\nu$-gap \cite{vinnicombe1993frequency,vinnicombe2000uncertainty} have been \textcolor{black}{studied intensively}. In this paper, we adopt the gap metric \textcolor{black}{as our main analysis tool}. Since the gap metric and its variants are topologically equivalent on the class of FDLTI systems, most of the results in this paper hold true for the $\nu$-gap and the pointwise gap with similar arguments.
\textcolor{black}{As for the non-ideal two-port communication channels}, we model their transmission matrices as
$\bm{T} = \bm{I} + \bm{\Delta}$, where $\bm{\Delta}$ is a bounded nonlinear operator. It is noteworthy that the NCS framework introduced in this paper can also accommodate certain nonlinear plants and controllers. In particular, a nonlinear plant or controller may be modelled as the interconnection of an FDLTI system and a nonlinear communication channel. For the ease of subsequent presentation, however, we adopt the convention of calling the two FDLTI components in a two-port NCS the plant and controller. \textcolor{black}{While it is well known that an uncertain system is usually represented by a fixed linear fractional transformation (LFT) of its uncertain component \cite{kimura1996chain,zhou1998essentials}, our two-port uncertainty model can be described by an uncertain LFT of a fixed component. }

The formulation of a two-port NCS is motivated by the
application scenario of stabilizing a feedback system where the plant and controller do not enjoy an ideal communication setup and their input-output
signals can only be sent through communication networks with several relays, as in, for example, teleoperation systems \cite{anderson1989bilateral},
satellite networks \cite{Alagoz2011POI}, wireless sensor networks \cite{Kumar2014CST} and so on. Moreover, each sub-system between two neighbouring
relays, which represents a communication channel, may introduce not only multiplicative distortions on the transmitted signal itself but also additive
interferences caused by the signal in the reverse direction, which usually arises in a bidirectional wireless network subject to channel fading \cite{tse2005fundamentals}
or under malicious attacks \cite{wu2007survey}.

The theory of two-port networks has been around for decades and served different purposes. \textcolor{black}{Historically, two-port networks were first introduced in electrical circuits theory \cite{choma1985electrical}.} Later on they were utilized to represent FDLTI systems in the
chain-scattering formalism \cite{kimura1996chain}, which is essentially a two-port
network.
\textcolor{black}{Such representations have also been used for studying feedback robustness from the perspective of the $\nu$-gap metric~\cite{Baskes2007CDC,Baskes2008CDC,Khong2013TAC}.}  Recently,
approaches based on the two-port network to modeling communication channels in a networked feedback system are studied in \cite{gu2011cdc,di2017cdcNonlinear,di2020tac}. 

The main contribution of this study is a clean result for concluding the finite-gain stability of an NCS with different types and multiple sources of uncertainties. \textcolor{black}{Technically speaking, a necessary and sufficient robust stability condition is obtained for FDLTI systems subject to simultaneous nonlinear perturbations.} While $\mu$ analysis \cite{zhou1998essentials} is known to be a general approach to deal with robust stabilization problems with structured uncertainties, it is computationally unattractive, and inadequate when it comes to modeling open-loop unstable dynamics as well as simultaneous uncertainties of different types within an NCS. By contrast, we approach the networked robust stabilization problem by taking advantage of the special two-port structures and making use of geometric
insights on \textcolor{black}{closed-loop} stability. 
\textcolor{black}{With the analysis from the geometric perspective,} we give a concise necessary and sufficient robust stability condition for the two-port NCS with nonlinear perturbations. Based on the condition, a special robust stability margin of the NCS is introduced in terms of the Gang of Four transfer matrix \cite{Astrom2008}.
\textcolor{black}{As for the synthesis of an optimally robust controller in the sense that the stability margin is maximized, it then suffices to solve a one-block $\mathcal{H}_\infty$ optimization problem \cite{georgiou1990optimal}. It is worth noting that similar geometric approaches to analyzing robust stability of nonlinear feedback systems have been developed from different perspectives; e.g., see \cite{teel1996TAC} for uncertainty with conic interpretation and \cite{megretski1997system,cantoni2012robustness}  for uncertainty subject to integral quadratic constraints.}

\textcolor{black}{There have been relevant works on robust stabilization of NCSs with special architectures and various uncertainty; see \cite{di2020tac} for a detailed introduction of literature. A previous study by the authors in \cite{di2017cdcNonlinear} considers a two-port NCS involving only nonlinear channel uncertainties under a rather strong assumption. This study differs from or generalizes the previous results in that it handles the model uncertainty and the nonlinear perturbations within two-port communication channels simultaneously. It allows for the modeling of a larger class of uncertain dynamics in NCSs with weak assumptions on nonlinearity, as demonstrated by a detailed example presented later in the paper.}

The rest of the paper is organized as follows. First in Section \ref{sec:pre}, we introduce open-loop \& closed-loop systems,  gap-type model uncertainties, and a preliminary robust stability result. Then in Section \ref{sec:network}, we present our main result of the robust stability condition for two-port NCSs. \textcolor{black}{In Section \ref{sec:simu}, a two-port NCS involving saturators and communication delay is simulated to demonstrate the efficacy of our result. The proof of the main theorem is provided in Section~\ref{sec:derivation}.} In Section \ref{sec:conc}, we conclude this study and \textcolor{black}{point out possible directions for future research}.

\section{Preliminaries}\label{sec:pre}
\subsection{Open-Loop Stability}


Let $\mathbb{F} = \mathbb{R}$ or $\mathbb{C}$ be the real or complex field,  and $\mathbb{F}^n$ be the linear space of $n$-tuples of $\mathbb{F}$ over the field $\mathbb{F}$. For $x \in \mathbb{F}^n$, its Euclidean norm is denoted by $|x|$. For a complex number $s\in\mathbb{C}$, its real and imaginary parts are denoted by $\Re\,s$ and $\Im\,s$, respectively.

Denote by $\mathcal{H}_\infty^{p\times m}$
the Hardy $\infty$-space of functions that are holomorphic and uniformly bounded on the right-half complex plane. This space is equipped with the $\mathcal{H}_\infty$ norm
$$\|G\|_\infty:=\sup_{\text{\rm Re}\,s>0}\bar{\sigma}(G(s))$$
for $G\in\mathcal{H}_\infty^{p\times m}$, where $\bar{\sigma}(\cdot)$ denotes the largest singular value of a complex matrix. An FDLTI system with transfer matrix $G$ is said to be stable if $G\in\mathcal{H}_\infty^{p\times m}$. Denote by $\mathcal{P}^{p\times m}$ the set of all real rational proper transfer matrices, and by $\mathcal{RH}_\infty^{p\times m}$ the set of all real rational members in $\mathcal{H}_\infty^{p\times m}$.

Denote the set of all causal energy-bounded signals by
$$\mathcal{L}_2^n
= \left\{u : [0, \infty) \to \mathbb{R}^n:~ \|u\|_2^2 := \int_0^\infty |u(t)|^2 \, dt < \infty\right\}.$$
For $T\geq 0$, define the truncation operator $\bm{\Gamma}_{T}$ on all signals $u : [0, \infty) \to \mathbb{R}^n$ by
$$(\bm{\Gamma}_{T} u)(t) = \left\{
\begin{array}{ll}
u(t), & \hbox{$0\leq t\leq{T}$;} \\
0, & \hbox{otherwise.}
\end{array}
\right.
$$
Denote the extended $\mathcal{L}_2$ space \cite[Chapter~2]{Willems1971nonlinear}, \cite[Chapter~8]{feintuch1982system} by
$$\mathcal{L}^n_{2e}:=\left\{u :[0, \infty) \to \mathbb{R}^n:~ \bm{\Gamma}_{T} u\in\mathcal{L}^n_2,~\forall~T>0\right\}.$$
It is noteworthy that $\mathcal{L}^n_{2e}$ is the completion of $\mathcal{L}^n_2$ with respect to the resolution topology \cite[Chapter~8]{feintuch1982system} defined via the family of seminorms $$\{\|\bmG_T(\cdot)\|_2:~T\in[0,\infty)\}.$$
A nonlinear system is represented by an operator
$$\bm{P}: ~\mathcal{L}_{2e}^m  \to  \mathcal{L}_{2e}^p.$$
We define the $\mathcal{L}_2$ domain of $\bmP$ as the set of all its input signals in $\mathcal{L}^m_2$ such that the output signals are in $\mathcal{L}^p_2$, i.e.,
$$\mathcal{D}(\bm{P}) := \{u \in \mathcal{L}^m_2 :~\bm{P} u \in \mathcal{L}^p_2\},$$
and correspondingly its $\mathcal{L}_2$ range as $\mathcal{R}(\bm{P}):=\bmP\mathcal{D}(\bm{P})$.

A physical system should
additionally be causal \cite[Chapters~2 and 4]{Willems1971nonlinear}, \cite[Chapter~6]{Vidyasagar1993nonlinear}, which is defined as follows.
\begin{definition}\label{def:causality}
A system $\bm{P}$ is said to be causal if for all ${T}> 0$ and $u_1,u_2 \in \mathcal{L}^m_{2e}$,
$$u_1(t)=  u_2(t),~t\in[0,T] \Rightarrow \bm{\Gamma}_{T} \bm{P} u_1 = \bm{\Gamma}_{T} \bm{P} u_2.$$
\textcolor{black}{It is said to be strongly causal if it is causal and if for all ${T}> 0$, $\epsilon>0$, and $T'>0$, $T'\leq T$, there exists a real number $\Delta_T>0$ such that for any $u_1,u_2 \in \mathcal{L}^m_{2e}$,
\begin{multline*} u_1(t) =  u_2(t),~t\in[0,T'] \Rightarrow\\ \|\bm{\Gamma}_{T'+\Delta_T} (\bm{P} u_1 -\bmP u_2)\| \leq \epsilon\| \bm{\Gamma}_{T+\Delta_T}( u_1- u_2)\|.\end{multline*}}
\end{definition}
\textcolor{black}{Roughly speaking, a strongly causal system possesses an infinitesimal delay.}


Throughout this study, we assume every system $\bmP$ is causal and that it has zero output whenever the input is zero, i.e., $\bm{P} 0 = 0$. \textcolor{black}{When $\bmP: \mathcal{L}_{2e}^m  \to  \mathcal{L}_{2e}^p$ is an FDLTI system, its restriction $\bmP|_{\mathcal{D}(\bm{P})}: \mathcal{D}(\bmP)\subset \mathcal{L}_{2}^m  \to  \mathcal{L}_{2}^p$ is equivalent, via the Fourier transform, to multiplication by a real rational proper transfer matrix in the frequency domain. On the other hand, an FDLTI system represented by a (possibly unbounded) linear operator $\bmP:~\mathcal{D}(\bmP)\to\mathcal{L}^p_2$ can be uniquely and causally extended to an operator mapping from $\mathcal{L}^m_{2e}$ to $\mathcal{L}^p_{2e}$ \cite[Proposition~11]{georgiou1993graphs}. Henceforth, we use $P\in\mathcal{P}^{p\times m}$ to denote the transfer matrix corresponding to such a $\bmP$, and we do not distinguish between an FDLTI system and its transfer matrix when the context is clear.}

The finite-gain stability of a system is defined as follows \cite[Chapter~6]{Vidyasagar1993nonlinear}.

\begin{definition}\label{def:finite_gain_stable}
A system $\bm{P}$ is said to be (finite-gain) stable if there exists $\alpha>0$ such that
\begin{align}\label{eq:def_finite_gain_stable}
\|\bmG_T\bmP u\|_2\leq \alpha\|\bmG_T u\|_2,~\forall~T\geq 0,~u\in\mathcal{L}_{2e}^m.
\end{align}
\end{definition}
\vspace{-10pt}
The following lemma is a direct consequence of \cite[Proposition~1.2.3]{arjan2017L2Gain}.
\begin{lemma}\label{lem:finite_gain_stability_def}
A system $\bmP$ is finite-gain stable if and only if $\mathcal{D}(\bm{P})=\mathcal{L}^m_2$ and
\begin{align*}
\|\bm{P}\| := \sup_{0 \neq x\in \mathcal{L}^m_2}\frac{\|\bm{P}x\|_2}{\|x\|_2}< \infty.
\end{align*}
When this is the case,
$$\|\bmP\|= \sup_{\substack{x\in\mathcal{L}_{2e}^m, T>0\\ \|\bm{\Gamma}_T x\|_2\neq 0}}\frac{\|\bm{\Gamma}_T \bmP x\|_2}{\|\bm{\Gamma}_T x\|_2}.$$
\end{lemma}
\vspace{-15pt}
\textcolor{black}{The incremental stability (or Lipschitz continuity) of a system is defined as follows \cite[Chapter~2]{Willems1971nonlinear}.
\begin{definition}\label{def:incremental_stable}
A system $\bmP$ is said to be incrementally stable if there exists a (Lipschitz) constant $L>0$ such that
$$\|\bmG_T(\bmP x-\bmP y)\|_2\leq L\|\bmG_T(x-y)\|_2,~\forall~T\geq 0,~x,y\in\mathcal{L}_{2e}^m.$$
\end{definition}}
\vspace{-10pt}
Clearly, the incremental stability of a system $\bmP$ implies its finite-gain stability since $\bmP 0=0$.


\subsection{Closed-Loop Stability}
Consider a standard closed-loop system $\nlcls$ as illustrated in Fig.~\ref{figloop} with plant $\bm{P}:  \mathcal{L}_{2e}^m \to \mathcal{L}_{2e}^p$
and controller $\bm{C}:  \mathcal{L}_{2e}^p \to \mathcal{L}_{2e}^m$. In the following, the superscripts may be omitted when the dimensions are clear from the context.

The graph of $\bm{P}$ is defined as
$$\mathcal{G}_{\bm{P}} = \begin{bmatrix}
\bm{I} \\
\bm{P} \\
\end{bmatrix}\mathcal{L}_{2e},$$
and similarly the inverse graph of $\bm{C}$ is defined as
$$\mathcal{G}'_{\bm{C}} = \begin{bmatrix}
\bm{C} \\
\bm{I} \\
\end{bmatrix}\mathcal{L}_{2e}.$$
The $\mathcal{L}_2$ graphs of $\bmP$ and $\bmC$ are defined as
$$\mathcal{G}^2_{\bm{P}}:=\mathcal{G}_{\bm{P}}\cap\mathcal{L}_2~~\text{and}~~\mathcal{G}'^2_{\bm{C}}:=\mathcal{G}'_{\bm{C}}\cap\mathcal{L}_2,$$ respectively, both of which are assumed to be closed throughout this paper. When $P$ and $C$ are FDLTI, $\mathcal{G}^2_P$ and $\mathcal{G}'^2_C$ are closed subspaces in $\mathcal{L}_2$.

It can be seen in \cite{Willems1971nonlinear,Vidyasagar1993nonlinear,SeiZhen_AUCC13} that various versions of feedback well-posedness may be stipulated based on different signal spaces and
causality requirements. In this study, we adopt the well-posedness definition from \cite{Willems1971nonlinear,Vidyasagar1993nonlinear}.
\begin{definition}\label{def:wellposed}
	The closed-loop system $\nlcls$ in Fig.~\ref{figloop} is said to be well-posed~if
	\begin{align*}
	\bm{F}_{\bm{P},\bm{C}}:&~\mathcal{L}_{2e}^m\times \mathcal{L}_{2e}^p \to \mathcal{L}_{2e}^{m+p}\\
	&=\begin{bmatrix}
	u_1 \\
	y_2 \\
	\end{bmatrix} \mapsto \begin{bmatrix}
	d_1 \\
	d_2 \\
	\end{bmatrix}= \begin{bmatrix}
	\bm{I} & -\bm{C}\\
	-\bm{P} & \bm{I} \\
	\end{bmatrix}
	\end{align*}
	is causally invertible on $\mathcal{L}_{2e}^{m+p}$.
\end{definition}
\textcolor{black}{Throughout the paper, well-posedness will always be assumed for the nominal as well as for all perturbed closed-loop systems.} Correspondingly, closed-loop stability is defined as follows.
\begin{definition}\label{def:closed_stable}
	A well-posed closed-loop system $\nlcls$ is (finite-gain) stable if $\bm{F}^{-1}_{\bm{P},\bm{C}}$ is finite-gain stable.
\end{definition}
%

When $\nlcls$ is well-posed, a pair of parallel projection operators~\cite{DOYLE199379,georgiou1997robustness}, $\bm{\Pi}_{\mathcal{G}_{\bm{P}}\sslash\mathcal{G}'_{\bm{C}}}$ along $\mathcal{G}'_{\bm{C}}$ onto $\mathcal{G}_{\bm{P}}$ and $\bm{\Pi}_{\mathcal{G}'_{\bm{C}}\sslash\mathcal{G}_{\bm{P}}}$
	along $\mathcal{G}_{\bm{P}}$ onto $\mathcal{G}'_{\bm{C}}$, can be defined
	respectively as
\begin{equation}\label{eq:pi_pc}
\begin{aligned}
\bm{\Pi}_{\mathcal{G}_{\bm{P}}\sslash\mathcal{G}'_{\bm{C}}}&: \mathcal{L}_{2e}^{m+p} \to \mathcal{G}_{\bm{P}}\\
&= \begin{bmatrix}
d_1 \\
d_2 \\
\end{bmatrix} \mapsto
\begin{bmatrix}
u_1 \\
y_1 \\
\end{bmatrix}= \bm{F}^{-1}_{\bm{P},\bm{C}}+\begin{bmatrix}
\bm{0} & \bm{0} \\
\bm{0} & -\bm{I} \\
\end{bmatrix},
\end{aligned}
\end{equation}
\begin{equation}\label{eq:pi_cp}
\begin{aligned}
\bm{\Pi}_{\mathcal{G}'_{\bm{C}}\sslash\mathcal{G}_{\bm{P}}}&: \mathcal{L}_{2e}^{m+p} \to \mathcal{G}'_{\bm{C}}\\
&=\begin{bmatrix}
d_1 \\
d_2 \\
\end{bmatrix} \mapsto
\begin{bmatrix}
u_2 \\
y_2 \\
\end{bmatrix}= \bm{F}^{-1}_{\bm{P},\bm{C}}+\begin{bmatrix}
-\bm{I} & \bm{0} \\
\bm{0} & \bm{0} \\
\end{bmatrix}.
\end{aligned}
\end{equation}
It follows that every $w \in \mathcal{L}_{2e}^{m+p}$ has a unique decomposition $w=u+v$ with $u=\bm{\Pi}_{\mathcal{G}_{\bm{P}}\sslash\mathcal{G}'_{\bm{C}}}w \in \mathcal{G}_{\bm{P}}$ and $v=\bm{\Pi}_{\mathcal{G}'_{\bm{C}}\sslash\mathcal{G}_{\bm{P}}}w \in \mathcal{G}'_{\bm{C}}$.

The next lemma bridges the closed-loop finite-gain stability and the boundedness of the above parallel projections \cite{DOYLE199379}.
\begin{lemma}\label{lem:pi_pc_stability}
	A well-posed closed-loop system $\nlcls$ is finite-gain stable if and only if  $\bm{\Pi}_{\mathcal{G}_{\bm{P}}\sslash\mathcal{G}'_{\bm{C}}}$ or $\bm{\Pi}_{\mathcal{G}'_{\bm{C}}\sslash\mathcal{G}_{\bm{P}}}$ is finite-gain stable.
\end{lemma}

For a finite-gain stable closed-loop system $\nlcls$, we can define its stability margin as
$\|\bm{\Pi}_{\mathcal{G}_{\bm{P}}\sslash\mathcal{G}'_{\bm{C}}}\|^{-1}$. It is shown in \cite{DOYLE199379} that if either $\bm{P}$ or $\bm{C}$ is linear, then
$\|\bm{\Pi}_{\mathcal{G}_{\bm{P}}\sslash\mathcal{G}'_{\bm{C}}}\|= \|\bm{\Pi}_{\mathcal{G}'_{\bm{C}}\sslash\mathcal{G}_{\bm{P}}}\|$. In particular, when $\lcls$ is FDLTI, the parallel projection operators reduce to the Gang of Four transfer matrix \cite{Astrom2008}, i.e.,
\vspace{-10pt}
\begin{align*}
\bm{\Pi}_{\mathcal{G}_{{P}}\sslash\mathcal{G}'_{{C}}} = \begin{bmatrix}
I \\
P \\
\end{bmatrix}
(I-CP)^{-1}
\begin{bmatrix}
I & -C
\end{bmatrix}=P\,\#\,C.
\end{align*}
Henceforth, for an FDLTI closed-loop system, $P\,\#\,C$ denotes both the closed-loop system itself and its Gang of Four transfer matrix.
%

\subsection{Gap Metric and Robust Stability}
A well-known tool for characterizing the model uncertainty in a system is the ``gap'' (or ``aperture'') and its variants \cite{zames1980proc,georgiou1988computation,georgiou1990optimal,qiu1992feedback,schumacher1992pointwise,qiu1992pointwise,vinnicombe1993frequency}. In what follows, we review some concepts related to the gap metric.

Let $\mathcal{X}$ and $\mathcal{Y}$ be two closed subspaces of a Hilbert space $\mathcal{L}$, and let $\Pi_{\mathcal{X}}$ and $\Pi_{\mathcal{Y}}$ be the orthogonal projections onto $\mathcal{X}$ and $\mathcal{Y}$, respectively. The gap metric between the two subspaces is defined as \cite[Chapter~2]{Kato1966PerturbationTheory}
\begin{align}\label{eq:gap}
\gamma (\mathcal{X},\mathcal{Y}):=\|\Pi_{\mathcal{X}}-\Pi_{\mathcal{Y}}\|.
\end{align}
The gap between FDLTI systems $P_1$ and $P_2$ is defined as the gap between their respective $\mathcal{L}_2$ graphs, i.e.,
\begin{align*}
\delta(P_1,P_2):=\gamma(\mathcal{G}^2_{P_1},\mathcal{G}^2_{P_2}).
\end{align*}
Given an FDLTI system $P\in\mathcal{P}$, denote the gap ball with center $P$ and radius $r \in [0,1)$ by
\begin{align}\label{eq:gapball}
 \mathcal{B}(P,r):=\left\{\tilde{P}\in\mathcal{P}:~\delta(P,\tilde{P})\leq r\right\}.
\end{align}
The following robust stability result, with the stability condition given in terms of an \textcolor{black}{``arcsine''} inequality, was obtained in \cite{qiu1992feedback}. We \textcolor{black}{state} it in the following lemma.
\begin{lemma}\label{lem:gaprobust}
Assume $\lcls\in\mathcal{RH}_\infty$. \textcolor{black}{For} $r_p, r_c \in [0,1)$, the perturbed system $\tilde{P}\,\#\,\tilde{C}$ is stable for all $\tilde{P} \in \mathcal{B}(P, r_p)$ and $\tilde{C} \in \mathcal{B}(C, r_c)$ if and only if
\begin{align}\label{eq:arcsin_standard}
\arcsin r_p + \arcsin r_c < \arcsin \| P\,\#\,C\|_\infty^{-1}.
\end{align}
\end{lemma}
\vspace{-10pt}
Lemma~\ref{lem:gaprobust} precisely quantifies the largest simultaneous gap-type uncertainties in the plant and controller, which the feedback system in Fig.~\ref{figloop} can tolerate while its stability is maintained. 
Naturally, the value $\|\lcls\|_\infty^{-1}$ can be regarded as the stability margin of the closed-loop system $\lcls$, as is consistent with its definition in Section~2.2.

\textcolor{black}{The design of the optimal robust controller can be attributed to solving an $\mathcal{H}_\infty$ problem with respect to the Gang of Four transfer matrix. Specifically, the optimization problem is aimed at computing the following maximum stability margin:
\begin{align}\label{eq:Copt_original}
\max_C \left\{\|\lcls\|_\infty^{-1}:~  C~\text{stabilizes}~P\right\}.
\end{align}
This is a special $\mathcal{H}_\infty$ problem, and can be reduced to {the Nehari problem, namely, the one-block $\mathcal{H}_\infty$ model matching problem,} which has been well studied and neatly solved \cite{glover1989robust,georgiou1990optimal,McF1990design}.}
\section{Main Results: Robust Networked Stabilization}\label{sec:network}
In this section, we present our main result of the study, which is a necessary and sufficient robust stability condition for a two-port NCS. Before establishing the condition, we first introduce some concepts related to two-port networks, and then investigate the stability of a two-port NCS subject to simultaneous uncertainties.
\subsection{Two-Port Networks as Communication Channels}

\begin{figure}
    \centering
    \begin{subfigure}[b]{0.24\textwidth}
        \includegraphics[width=\textwidth]{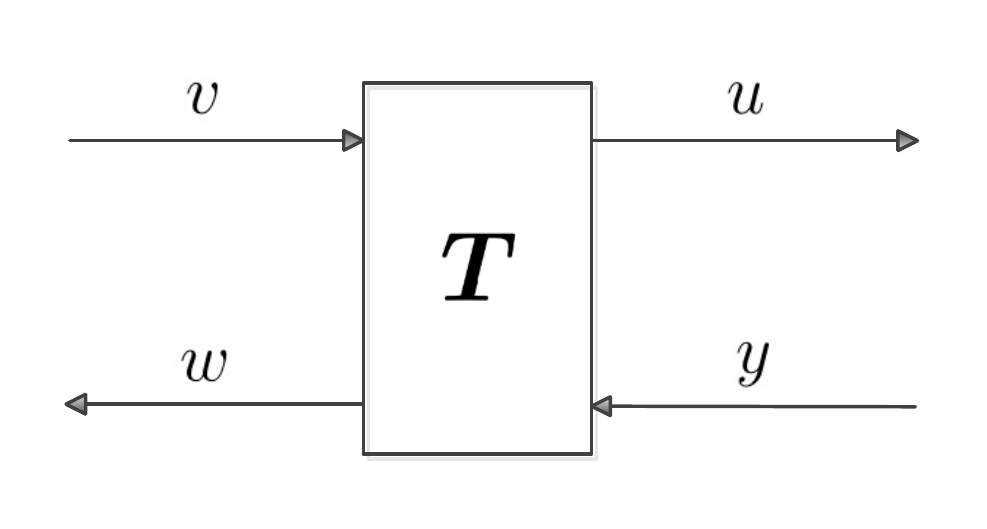}
        \caption{Single two-port network.}
        \label{figtwoport}
    \end{subfigure}
    ~ 
    \begin{subfigure}[b]{0.28\textwidth}
        \includegraphics[width=\textwidth]{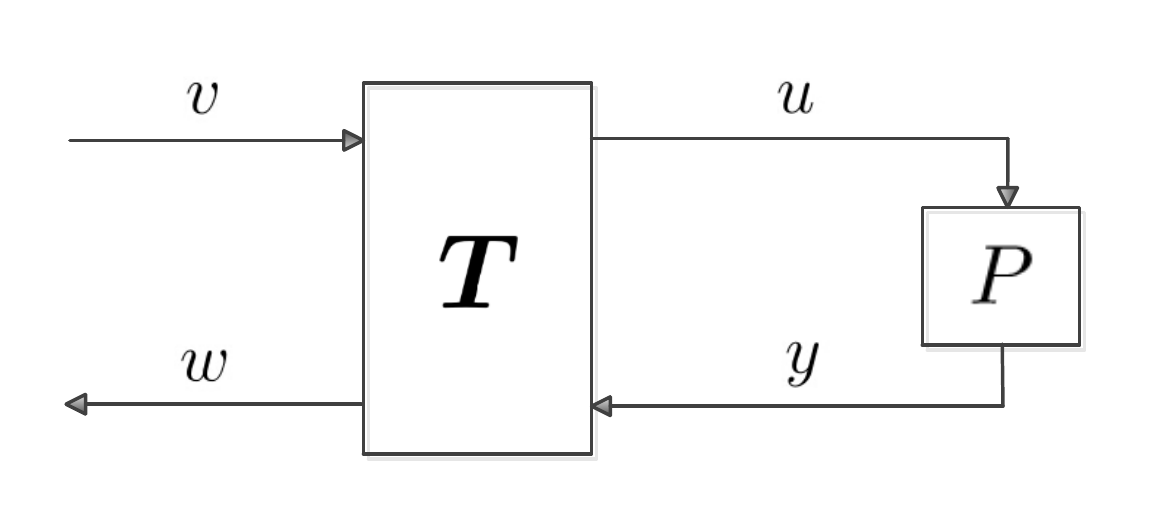}
        \caption{One-stage two-port connection.}
        \label{fig:oneblock}
    \end{subfigure}
    \begin{subfigure}[b]{0.31\textwidth}
        \includegraphics[width=\textwidth]{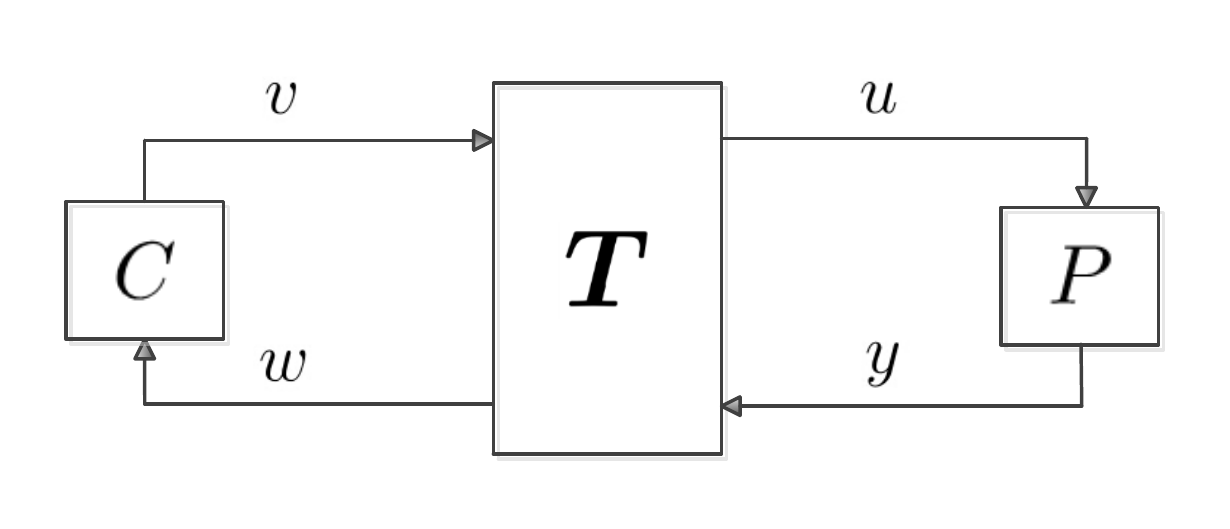}
        \caption{Two-port NCS with one channel.}
        \label{fig:oneblock_feedback}
    \end{subfigure}
    \caption{Two-port network $\bmT$. }\label{fig:two_port_of2}
    \vspace{-5pt}
\end{figure}

The use of two-port networks as a model of communication channels is adopted from \cite{gu2011cdc,di2020tac}. The network $\bmT$ in Fig.~\ref{fig:two_port_of2}(a) has two
external ports, with one port composed of $v$, $w$ and the other of $u$, $y$, and is hence called a two-port network. A two-port network $\mathbd{T}$
has various representations with respect to different parameters, such as impedance parameters, admittance parameters, hybrid parameters, transmission parameters and so on, among which we choose the transmission (parameter) representation to model a bidirectional communication channel. Define the transmission matrix $\bm{T}$ as
\begin{align}\label{eq:transmission}
\bm{T} = \begin{bmatrix}
\bm{T}_{11} & \bm{T}_{12} \\
\bm{T}_{21} & \bm{T}_{22} \\
\end{bmatrix}~\text{and} ~\begin{bmatrix}
     v \\
     w \\
   \end{bmatrix} =  \bm{T}\begin{bmatrix}
     u \\
     y \\
   \end{bmatrix}.
\end{align}
Henceforth, $\bm{T}$ stands for both the two-port network itself and its transmission representation for notational simplicity.

When the communication channel is perfect, i.e., communication takes place without distortion or interference, the transmission matrix is simply
$$\bm{T} = \begin{bmatrix}
    \bm{I}_m & \bm{0} \\
    \bm{0} & \bm{I}_p \\
    \end{bmatrix},$$
where $\bm{I}_n: \mathcal{L}^n_{2e} \to \mathcal{L}^n_{2e}$ is the identity operator.
If the bidirectional channel admits both distortions and interferences, we elucidate below that a good modeling approach involves the transmission matrix of the form
$$\bm{T} = \bm{I}+\bm{\Delta}= \begin{bmatrix}
\bm{I}_m+\bm{\Delta}_\div & \bm{\Delta}_- \\
\bm{\Delta}_+ & \bm{I}_p+\bm{\Delta}_\times \\
\end{bmatrix},$$
 where $$\bm{\Delta} = \begin{bmatrix}
\bm{\Delta}_\div & \bm{\Delta}_- \\
\bm{\Delta}_+ & \bm{\Delta}_\times \\
\end{bmatrix}$$
is finite-gain stable with $\|\bm{\Delta}\|<1$. \textcolor{black}{We also assume that $\bmD$ is strongly causal as introduced in Definition~\ref{def:causality}.} In this case, $\bmI\,\#\,(-\bmD)$ is well-posed, and it follows from the nonlinear small-gain theorem \cite[Chapter~3]{desoer1975feedback} that $\bm{T}^{-1}$ is causal and finite-gain stable. The four-block operator matrix $\bm{\Delta}$ is called the (nonlinear) uncertainty quartet \cite{di2018bookchapter}.

\begin{figure}
  \centering
  \includegraphics[scale=0.46]{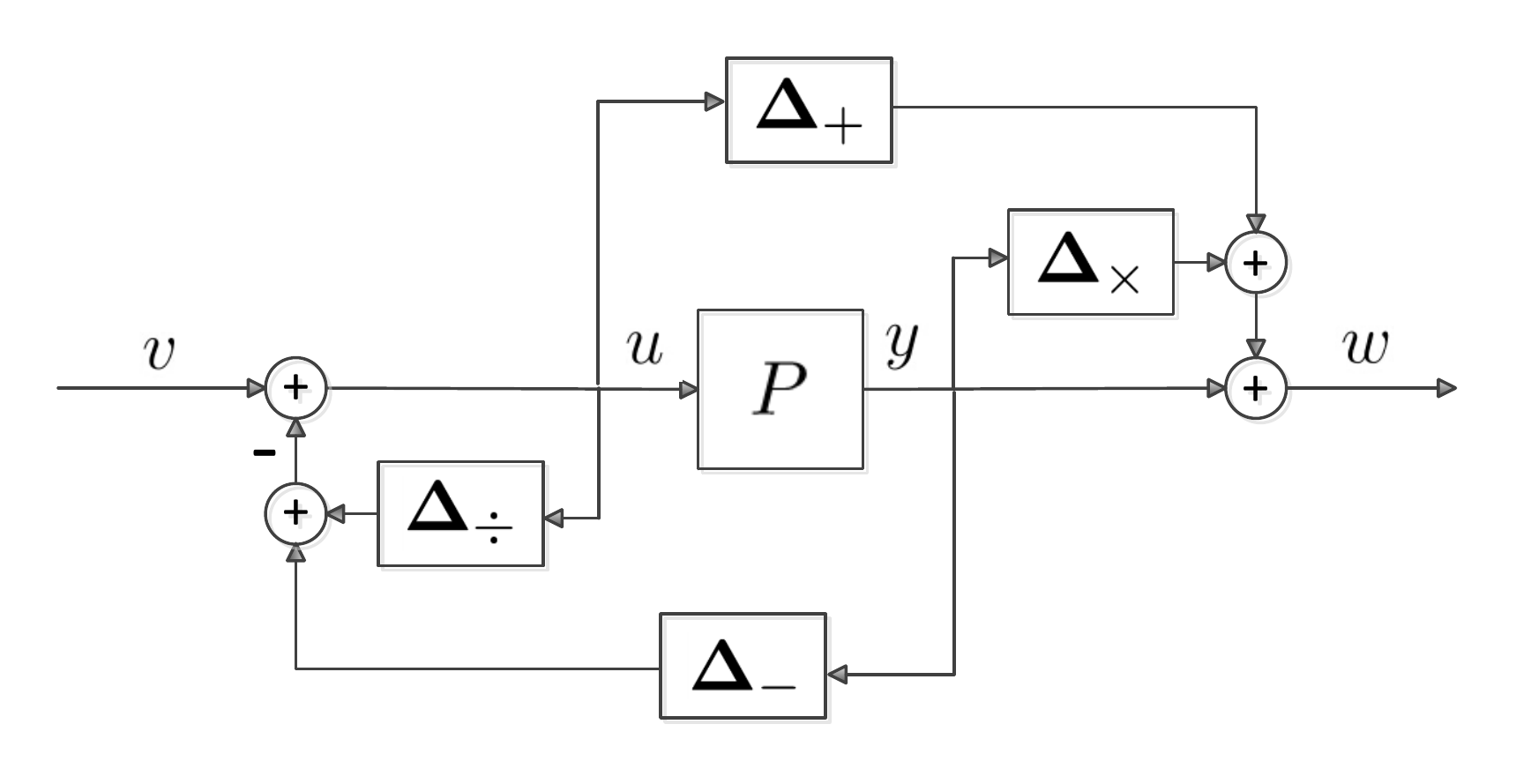}\\
  \caption{Plant with the uncertainty quartet.}\label{fig:LFT_two-port}
\end{figure}

\textcolor{black}{Fig.~\ref{fig:two_port_of2}(b) describes a two-port network $\bmT=\bmI+\bmD$ connected to an FDLTI system $P$.} One way to analyze how the uncertainties influence the nominal system is via the transformation of the diagram into the one in Fig.~\ref{fig:LFT_two-port}. In this \textcolor{black}{case}, we can interpret the members in the uncertainty quartet with their respective physical meanings~\cite{halsey2005analysis}, namely, the divisive ($\div$), the subtractive ($-$), the additive ($+$) and the multiplicative ($\times$) uncertainty. 
\begin{remark}\label{rmk:causal}
\textcolor{black}{It follows from the strong causality of $\bmD_-$ and $\bmD_\div$ that the feedback loop in Fig.~\ref{fig:LFT_two-port} is well-posed. As a result, the perturbed system $v\mapsto w$ is well defined and causal. }
\end{remark}

The diagonal terms $\bmD_{\div},\bmD_{\times}$ and the off-diagonal terms $\bmD_{-},\bmD_{+}$ model two types of perturbations. The diagonal terms model multiplicative nonlinear distortions of the transmitted signals, mostly due to signal attenuations in the fading channel. The off-diagonal terms represent additive interferences caused by the reverse signals, which occurs mostly in bidirectional communication channels. 

\begin{figure}
  \centering
  \includegraphics[scale=0.6]{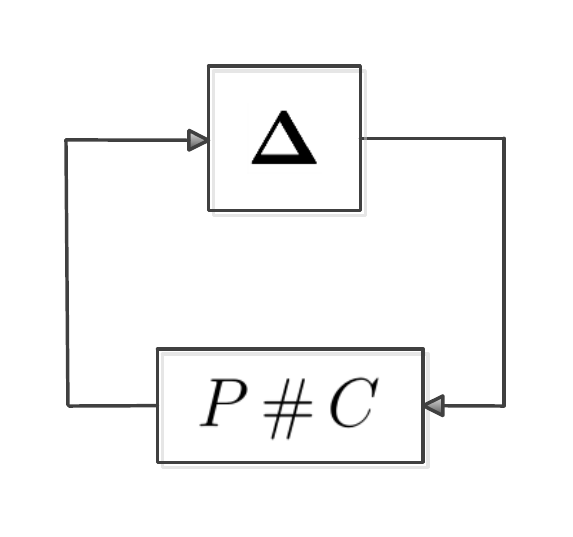}\\
  \caption{\textcolor{black}{Standard closed-loop system reformulated from an equivalent one-stage two-port NCS.}}\label{fig:Structured_onestage}
\end{figure}
Consider the situation where we close the loop in Fig.~3(b) with an FDLTI controller $C$, which stabilizes $P$, as in Fig.~3(c). Following the derivation in \cite{gu2011cdc}, we can separate the uncertainty quartet $\bmD$ and the nominal closed-loop system $\lcls$ to obtain an equivalent closed-loop system $(\lcls)\,\#\,\bmD$ as shown in Fig.~\ref{fig:Structured_onestage}. The robust stability of this system \textcolor{black}{can be analyzed using} the nonlinear small-gain theorem \cite[Chapter 3]{desoer1975feedback}, \textcolor{black}{resulting in the following stability condition}.
\textcolor{black}{\begin{lemma}\label{lem:two-port-smallgain}
Assume $\lcls\in\mathcal{RH}_\infty$. For $r \in [0,1)$, the two-port NCS in Fig.~3(c) is stable for all $\bmD$ with $\|\bmD\| \leq r$ if and only if
\begin{align}\label{eq:small_gain}
r < \|P\,\#\,C\|^{-1}_\infty.
\end{align}
\end{lemma}}
Here, the stability margin $\|P\,\#\,C\|^{-1}_\infty$ comes into the picture again, \textcolor{black}{as it appears} in Lemma~\ref{lem:gaprobust} for gap-type uncertainties. This motivates us to \textcolor{black}{obtain} a unified robust stability condition with combined gap-type model uncertainties and two-port uncertainty quartets, which \textcolor{black}{is the object of study in what follows}.

\subsection{Graph Analysis on Cascaded Two-Port NCS}\label{subsec:NCS with Two-Port}
\begin{figure}
  \centering
  \includegraphics[scale=0.45]{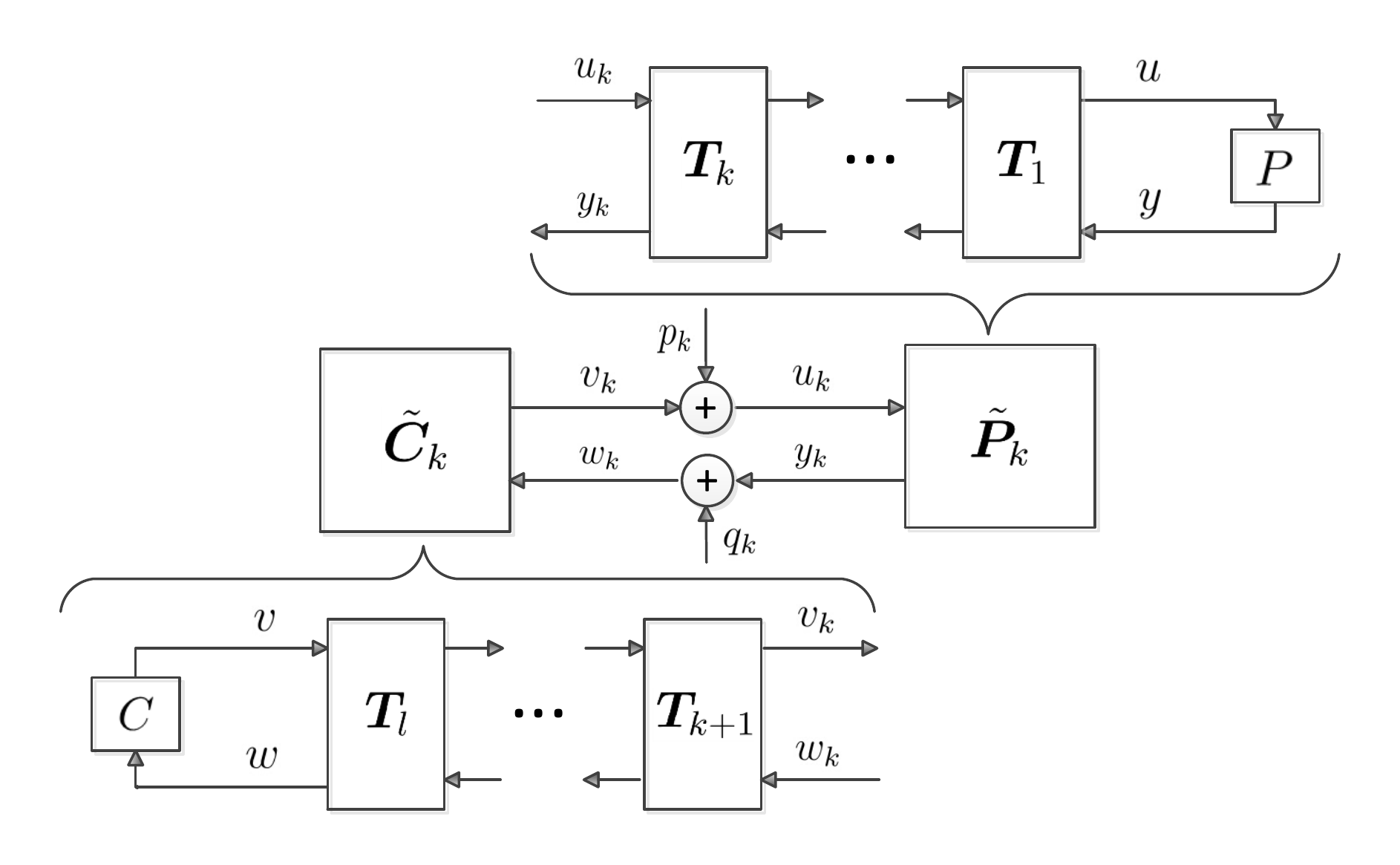}\\
  \caption{Perturbed plant and controller representations.}\label{fig:equiv_con_pla}
  \vspace{-5pt}
\end{figure}

In order to acquire a better understanding of a two-port NCS, we establish in the following the connection between a two-port NCS and its equivalent closed-loop systems by investigating the graphs of systems.
As in Fig.~\ref{fig:equiv_con_pla}, for each integer $k \in [1,l]$, the $k$-th stage two-port network, which admits a \textcolor{black}{strongly causal and} incrementally stable nonlinear uncertainty $\bm{\Delta}_k$, is represented by the transmission matrix $\bm{T}_k =
\bm{I}+\bm{\Delta}_k$.  We can associate the first $k$ stages of the cascaded two-port networks with plant $P\in\mathcal{P}^{p\times m}$, and the remaining $l-k$ stages with the controller $C\in\mathcal{P}^{m\times p}$. Then the diagram in Fig.~\ref{fignetwork} is equivalently transformed into that in Fig.~\ref{fig:equiv_con_pla} to form a standard closed-loop system $\tilde{\bmP}_k\,\#\,\tilde{\bmC}_k$. 

With similar analysis in \cite[Section III.B]{di2020tac}, we view $P$ together with $\{\bmT_j\}_{j=1}^{k}$ as \textcolor{black}{a perturbed plant $\tilde{\bm{P}}_k=u_k\mapsto y_k$} with uncertainties $\{\bm{\Delta}_j\}_{j=1}^{k}$, and thereby obtain the
graph of $\tilde{\bm{P}}_k$ as
\begin{equation}\label{eq:G_Pk}
\begin{aligned}
\mathcal{G}_{\tilde{\bm{P}}_k} = (\bm{I}+\bm{\Delta}_{k})\cdots (\bm{I}+\bm{\Delta}_{1})\mathcal{G}_{P}.
\end{aligned}
\end{equation}
Similarly, we view $C$ together with $\{\bmT_j\}_{j=k+1}^{l}$ as a perturbed controller \textcolor{black}{$\tilde{\bm{C}}_k=w_k \mapsto v_k$} with uncertainties
$\{\bm{\Delta}_j\}_{j=k+1}^{l}$, and obtain the inverse graph of $\tilde{\bm{C}}_k$ as
\begin{equation}\label{eq:G_Ck}
\begin{aligned}
\mathcal{G}'_{\tilde{\bm{C}}_k} = (\bmI+\bm{\Delta}_{k+1})^{-1}\cdots (\bmI+\bm{\Delta}_{l})^{-1}\mathcal{G}'_{C}.
\end{aligned}
\end{equation}
For convenience, we include $k = 0$ and $k=l$ with the interpretation that the entire two-port network is grouped with controller $C$ for $k=0$, and with plant $P$ for $k=l$. \textcolor{black}{Similarly to the discussions in Remark~\ref{rmk:causal}, we obtain from the strong causality of $\bmD_k$, $k=1,2,\dots,l$, that $\tilde{\bmP}_k$ and $\tilde{\bmC}_k$, $k=0,1,\dots,l$, are well defined and causal.}



\subsection{Main Theorem: Robust Stability Condition}
With the perturbed plant and controller representations derived as above, next we extend the definition of the stability of the two-port NCS in
\cite{di2020tac} to the nonlinear setting.

As shown in Fig.~\ref{fig:equiv_con_pla}, we denote the $k$-th input pair of the NCS as $I_k := [p_k~q_k]^T$, the $k$-th output pair as $O_k:= [u_k~w_k]^T$ and
the vector collecting all outputs as $O := [O_1 ~O_2 ~\cdots ~O_l]^T$. By the feedback well-posedness assumption, the map from $I_k\in\mathcal{L}_{2e}$ to $O\in\mathcal{L}_{2e}$
is well defined, \textcolor{black}{which is denoted as a causal operator $\bm{A}_k=I_k\mapsto O$.}
\begin{definition}\label{def:NCS_stability}
The two-port NCS in Fig.~\ref{fig:equiv_con_pla} is said to be stable if $\bm{A}_k$ is finite-gain stable for every $k=0,1,\dots,l$.
\end{definition}
The following proposition further simplifies the stability condition of an NCS, whose proof is in Appendix
\begin{proposition}\label{prop:stability}
	The two-port NCS is finite-gain stable if and only if the equivalent closed-loop system $\tilde{\bm{P}}_k\,\#\,\tilde{\bm{C}}_k$ is finite-gain stable for every $k = 0,1,...,l$.
\end{proposition}
Given the definition of stability of NCSs, we are ready to present the main robust stability theorem involving simultaneous uncertainties in the two-port NCS.
Specifically, the communication channels are subject to nonlinear perturbations, and the plant and controller are subject to model uncertainties characterized by the gap metric. The proof of the theorem is provided in Section~\ref{sec:derivation}.

\begin{theorem}\label{thm:NL_main}
\textcolor{black}{Let $P$ or $C$ be strongly causal,} $\lcls\in\mathcal{RH}_\infty$ and $r_p,r_c,r_k \in [0,1)$.
The NCS in Fig.~\ref{fignetwork} is finite-gain stable for all $\tilde{P} \in \mathcal{B}(P, r_p), ~\tilde{C} \in \mathcal{B}(C, r_c)$, and \textcolor{black}{strongly causal and} incrementally stable $\bm{\Delta}_k$ with $\|\bm{\Delta}_k\| \leq r_k$, $k=1,2,\dots,l$, if and only if\vspace{-10pt}
\begin{multline}\label{eq:NL_two-port_arcsin}
\arcsin r_p + \arcsin r_c + \sum _{k=1}^{l} \arcsin r_k\\
 < \arcsin \|P\,\#\,C\|^{-1}_\infty.
\end{multline}
\end{theorem}
From the above theorem, it is clear that $\|P\,\#\,C\|^{-1}_\infty$ can be viewed as the robust stability margin of the two-port NCS. The larger the margin, the more robust the two-port NCS.
In addition, we know the robust stability margin $\|P\,\#\,C\|^{-1}_\infty$ is the same as that in a standard closed-loop system in the presence of gap-type uncertainties as in Lemma~\ref{lem:gaprobust} or of uncertainty quartets as in Lemma~\ref{lem:two-port-smallgain}. As a result, the controller synthesis problem of a two-port NCS can be solved \textcolor{black}{via the same $\mathcal{H}_\infty$ control problem as that in \eqref{eq:Copt_original}}. In addition, the optimal synthesis is independent of the structure of the communication channel between the plant and controller, such as the number of two-port connections and how the uncertainty bounds are distributed among all the channels.



%
\section{Simulation: An NCS with Saturator and Communication Delay}\label{sec:simu}

\textcolor{black}{Consider a nominal double integrator system}
\[
P(s) = \frac{1}{s^2},
\]
\textcolor{black}{which may model an ideal rigid body undergoing a forced linear motion. This system has been studied in various textbooks \cite{doyle1990feedback,qiutextbook}.
	The corresponding optimally robust controller $C$, maximizing the robust stability margin $\|P\,\#\,C\|_\infty^{-1}$, is given by}
$$C(s) = -\frac{(1+\sqrt{2})s + 1}{s+1+\sqrt{2}}.$$
{\color{black}The associated} optimally robust stability margin is {\color{black}obtained as} $$ \|P\,\#\,C\|_\infty^{-1} = \left(4+2\sqrt{2}\right)^{-1/2}.$$ 

\begin{figure}
	\centering
	\vspace{-15pt}
	\includegraphics[scale=0.62]{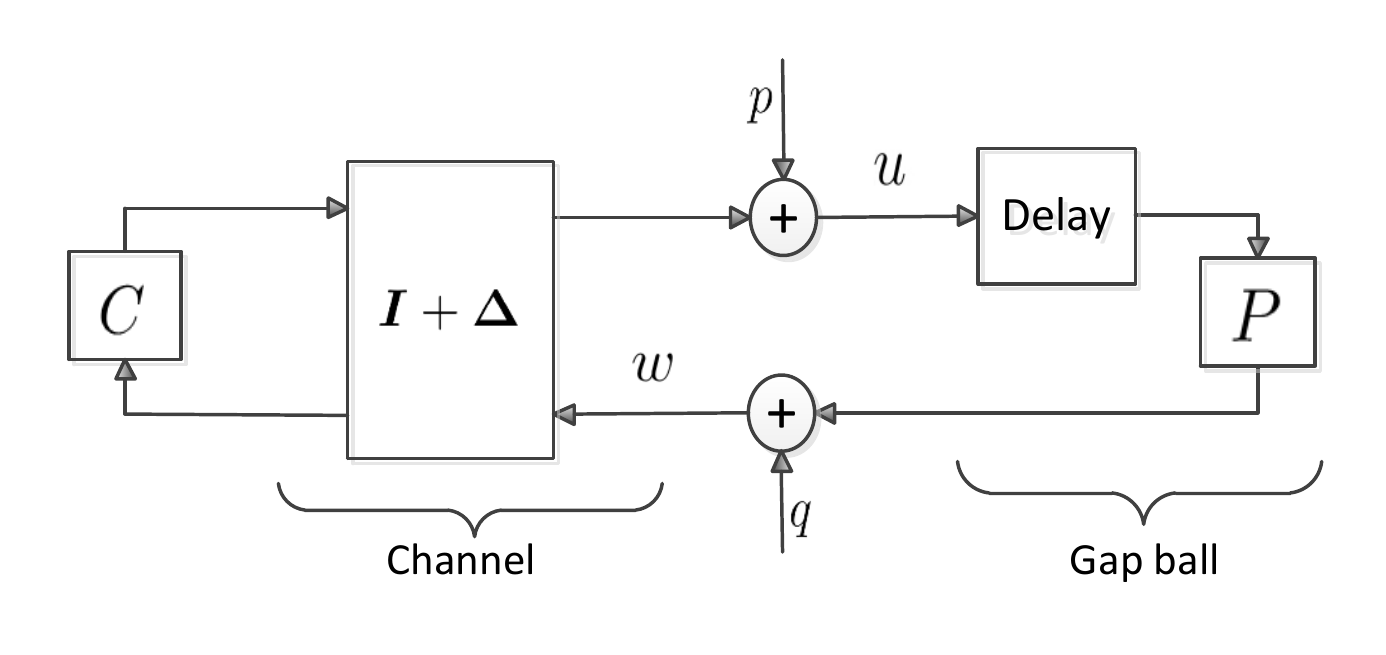}\\
	\vspace{-15pt}
	\caption{An NCS involving saturation and time delay. }\label{fig:simu_diagram}
\end{figure}
As shown in Fig.~\ref{fig:simu_diagram}, the channel is modelled by a two-port network involving saturation. In addition, uncertain communication delay is present at the input to the plant, which may be characterized using the gap metric.

In terms of the communication channel modelled by a two-port network, let its transmission matrix be $\bmT = \bmI + \bmD$, where the nonlinear uncertainty quartet is
$$\bm{\Delta} = \begin{bmatrix}
0 & -\frac{\sqrt{3}}{2}{r}\bm{\Lambda}g \\
0 & -\frac{1}{2}{r}g \\
\end{bmatrix}$$
with parameter ${r} \in (0,1)$, a standard saturator
$$\bm{\Lambda}:\mathbb{R}\to\mathbb{R}=x \mapsto \left\{
\begin{array}{ll}
x, & \hbox{$|x|\leq 1$} \\
1, & \hbox{$|x|>1$}
\end{array}
\right.
$$
\textcolor{black}{and a strictly proper transfer function $g=\alpha/(\alpha+s)$ for $\alpha>0$. Since $g$ is strictly proper, $\bmD$ is strongly causal.}
The saturator, located on an off-diagonal position of the uncertainty quartet $\bmD$, is part of the subtractive uncertainty, which may characterize the attenuation of the interferences within the two-port communication channel. \textcolor{black}{The gain of the uncertainty quartet is $\|\bmD\| = {r}$. We can verify, according to \cite[Chapter~4]{Willems1971nonlinear}, that its uniform instantaneous gain is less than $1$, and thus $\bmT$ is causally and stably invertible with its inverse given by}
$$\bmT^{-1}=\begin{bmatrix}
1 & \frac{\sqrt{3}}{2}{r}\bm{\Lambda}g(1-\frac{1}{2}{r})^{-1} \\
0 & (1-\frac{1}{2}{r}g)^{-1} \\
\end{bmatrix}.$$
\begin{figure}
	\centering
	\includegraphics[scale=0.4]{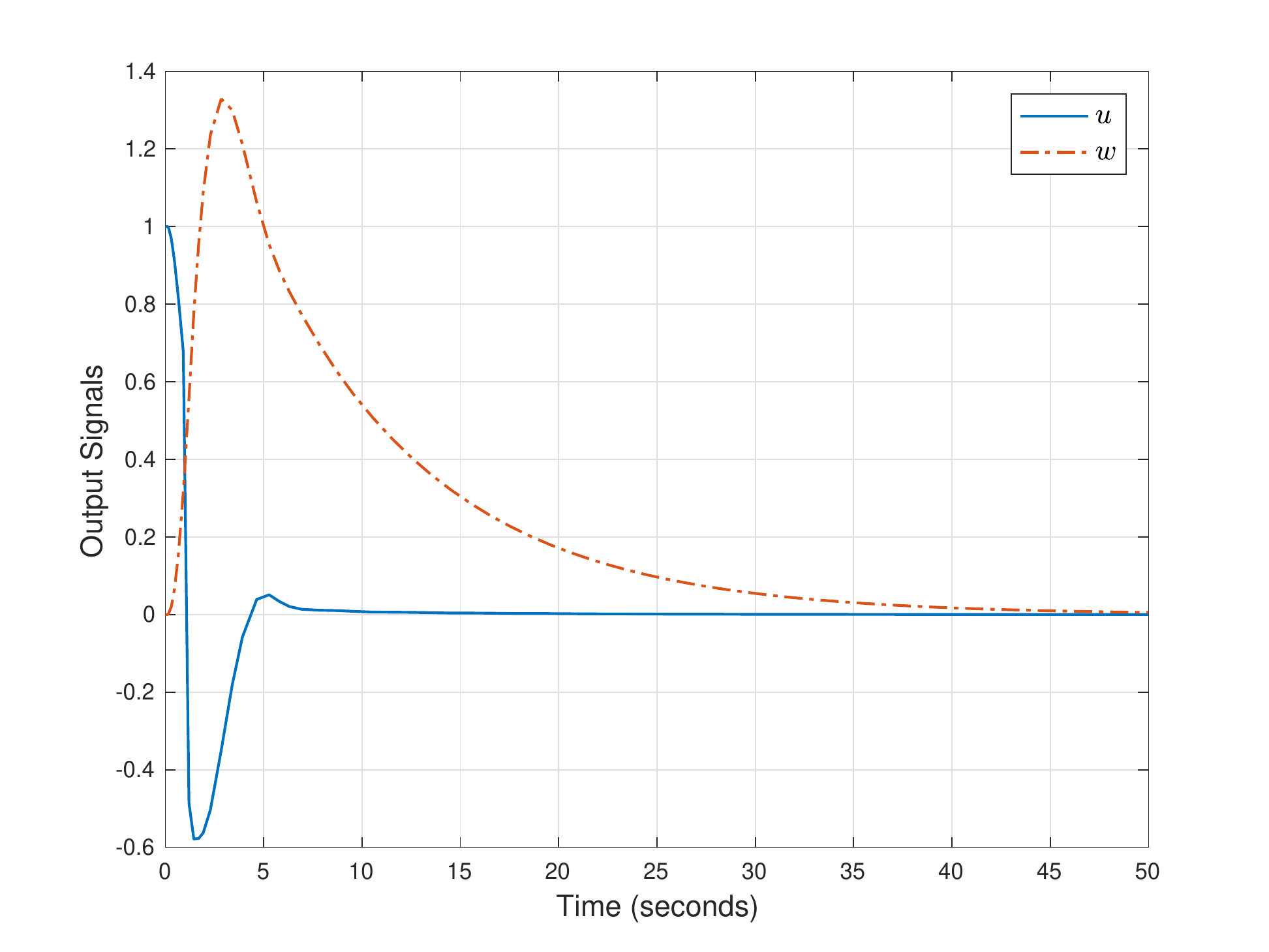}\\
	\caption{Output evolution when condition (\ref{eq:NL_two-port_arcsin}) holds.}\label{fig:NL_simu_stable}
\end{figure}

As for the time delay, we incorporate it into the side of the plant \cite{dym1995explicit}. 
The plant $P$ is {thus replaced by the delayed model}  $Pe^{-2\tau s}$ with $\tau>0$ being the time delay for one-way transmission.

\begin{figure}
	\centering
	\includegraphics[scale=0.42]{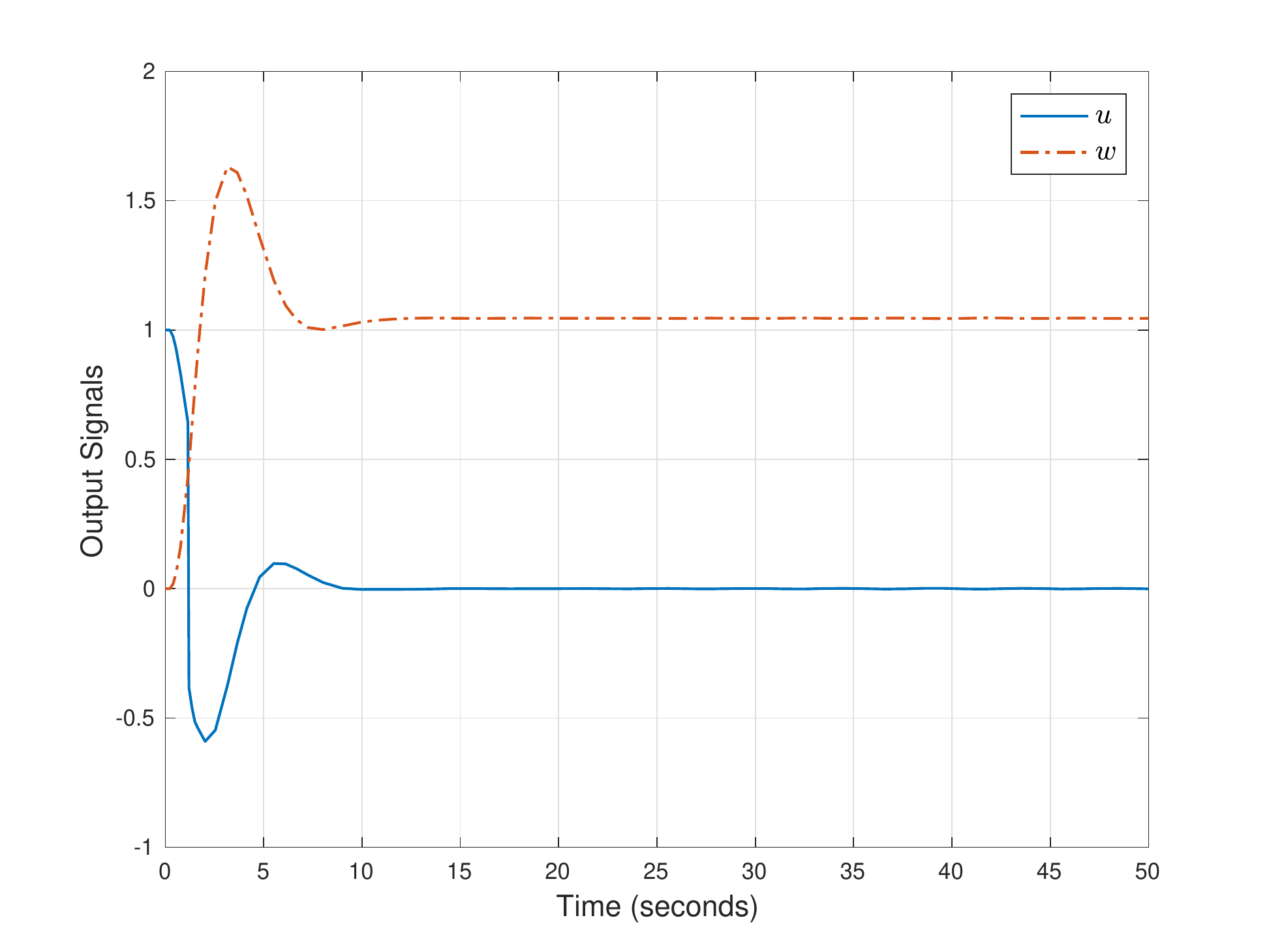}\\
	\caption{Output evolution when condition (\ref{eq:NL_two-port_arcsin}) is violated.}\label{fig:NL_simu_unstable}
\end{figure}
As an illustrative example, let $\alpha\gg 0$, ${r}=0.32$ and $\tau = 0.1 [\text{sec}]$. We have ${r}_2 = \max\{\delta_d,\delta_u\} = 0.18$ and ${r}_p = \delta(P,Pe^{-2\tau s})<0.0576$, as estimated by the Pad{\'e} approximation \cite{baker1996pade}. Hence,
\begin{multline*}
\arcsin {r}_1 + \arcsin {r}_2 + \arcsin {r}_p = 0.3834\\
< 0.3927 = \arcsin \|P\,\#\,C\|_\infty^{-1},
\end{multline*}
{\color{black}which marginally satisfies} the stability condition (\ref{eq:NL_two-port_arcsin}).
It can be seen from Fig.~\ref{fig:NL_simu_stable} that the two-port NCS is stable when stimulated by an impulse signal at $p$.
\textcolor{black}{Since the nonlinear perturbations in the two-port channels due to saturation and time-delay do not correspond to the worst-case scenario, stability of the NCS is maintained even when condition (\ref{eq:NL_two-port_arcsin}) is slightly violated. Nevertheless, we can observe that the system approaches instability as the values of the parameters increase. Indeed, when the parameters reach ${r} = 0.4$ and $\tau = 0.2 [\text{sec}]$, we have}
\begin{multline*}
\arcsin {r} + \arcsin {r}_p = 0.5259\\
> 0.3927= \arcsin \|P\,\#\,C\|_\infty^{-1},
\end{multline*}
 and the NCS is unstable as shown in Fig.~\ref{fig:NL_simu_unstable}. 

\section{Derivation of the Main Result}\label{sec:derivation}
This section is dedicated to the derivation of the main result --- Theorem~\ref{thm:NL_main}.

\subsection{Conelike Uncertainty Sets}
\begin{figure}
	\centering
	\includegraphics[width=.35\textwidth]{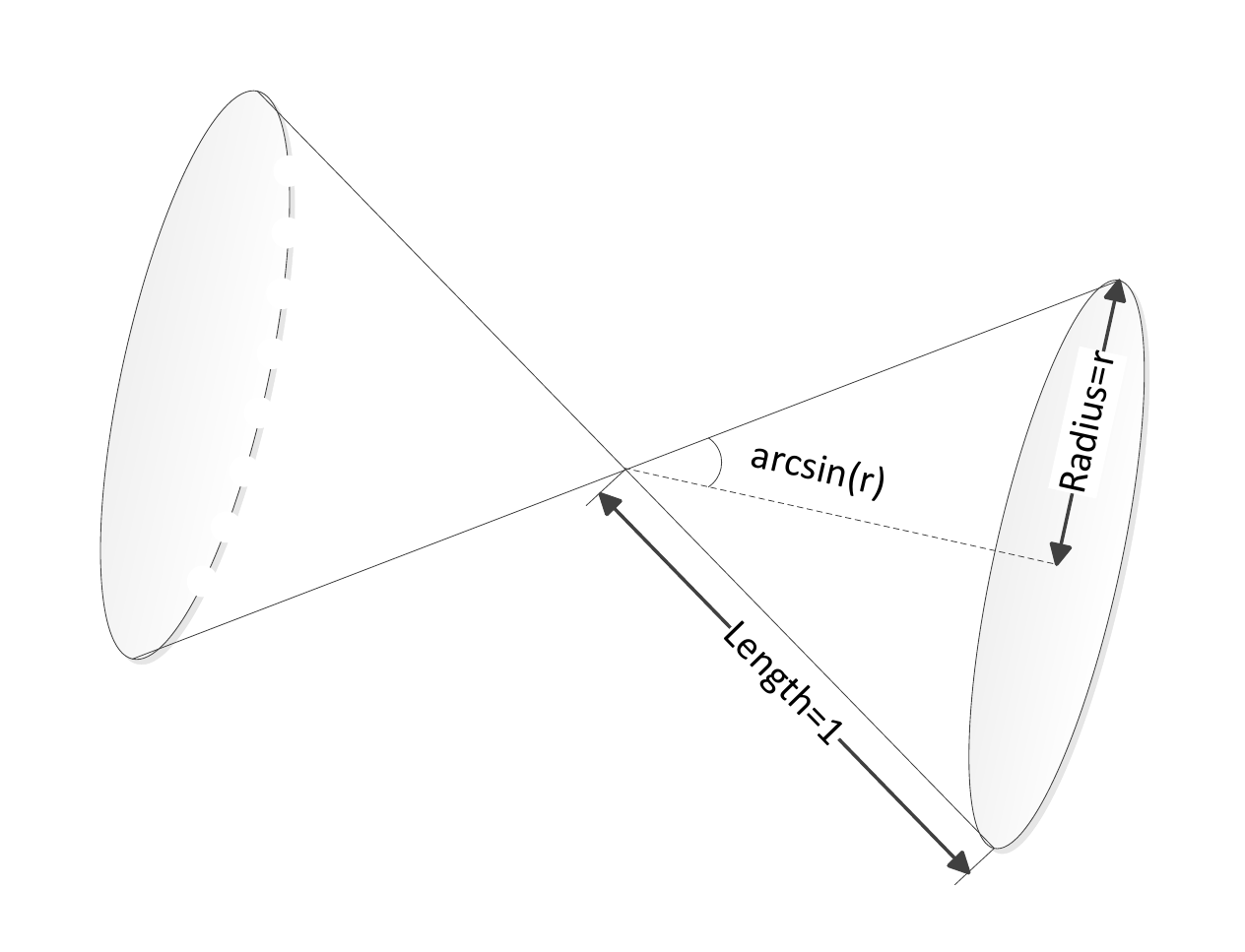}\\
	\caption{A conelike neighborhood in $\mathbb{R}^3$ with radius $r$. }\label{fig:double_cone}
\end{figure}
As for FDLTI systems, the gap metric and its variants have achieved great success in characterizing model uncertainties. In order to characterize nonlinear uncertainties in the spirit of the gap metric, we introduce the notion of conelike uncertainty sets as follows. Let $\mathcal{M}$ be a \textcolor{black}{closed set} in $\mathcal{L}_2$. Define
the conelike neighborhood centered at $\mathcal{M}$ as
\begin{align}\label{eq:conelike_neighboring_definition}
\hspace{-5pt}\mathcal{S}(\mathcal{M},r) \hspace{-2pt}:=\hspace{-2pt} \left\{v\in \mathcal{L}_2 :\inf_{0 \neq u\in \mathcal{M}}\hspace{-2pt} \frac{\|v-u\|_2}{\|u\|_2}\hspace{-1pt}\leq\hspace{-1pt} r\right\}\cup \{0\}.
\end{align}
If $\mathcal{M}$ is a one-dimensional subspace in $\mathbb{R}^3$, the set $\mathcal{S}(\mathcal{M},r)$ is simply a right circular double cone (twin cone) as shown in Fig.~\ref{fig:double_cone}. In the case where $\mathcal{M}$ is a one-dimensional subspace in $\mathbb{R}^2$, the set $\mathcal{S}(\mathcal{M},r)$ can be interpreted as \textcolor{black}{a doubly sector-bounded area \cite{zames1966sector}.} If $\mathcal{M}$ is the $\mathcal{L}_2$ graph of a certain FDLTI system, the set $\mathcal{S}(\mathcal{M},r)$ \textcolor{black}{can be viewed as} a closed double cone in $\mathcal{L}_2$, which provides some geometric intuition on the system uncertainties. 

Similarly to the conelike neighborhood defined in (\ref{eq:conelike_neighboring_definition}), the inverse conelike neighborhood is defined as follows:
$$\textcolor{black}{\tilde{\mathcal{S}}(\mathcal{M},r):= \left\{v\in \mathcal{L}_2 :~ \inf_{u\in \mathcal{M}} \frac{\|v-u\|_2}{\|v\|_2}\leq r\right\}\cup \{0\}.}$$

We have the following useful proposition describing the properties of the conelike neighborhoods. Based on the Hilbert space structure of $\mathcal{L}_2$, the acute angle between $x,y \in \mathcal{L}_2\setminus \{0\}$ is denoted by
$$\theta(x,y):=\arccos\frac{|\langle x,y\rangle|}{\|x\|_2\|y\|_2},$$
and let $\theta(x,y)=\infty$ if either of $x,y$ is zero almost everywhere for technical simplicity.
\begin{proposition}\label{prop:properties_conelike}
	Let $ r_0,r \in [0,1)$, $\mathcal{M}_0\subset\mathcal{L}_2$ be a closed subspace, and $\mathcal{M}=\mathcal{S}(\mathcal{M}_0,r_0)$. The following statements are true.
	\begin{enumerate}
		\item [(a)]  Let $ v \in \mathcal{L}_2\setminus \{0\}$. Then it follows that $v \in \mathcal{S}(\mathcal{M},r)$ if and only if $\alpha v \in \mathcal{S}(\mathcal{M},r)$ for all $\alpha \in \mathbb{R}$.
		\item [(b)] It holds that\vspace{-10pt}
		\begin{align*}
		&\mathcal{S}(\mathcal{M},r) = \tilde{\mathcal{S}}(\mathcal{M},r)\\
		&=\left\{v\in \mathcal{L}_2 : \inf_{u\in \mathcal{M}} \theta(u,v)\leq \arcsin r\right\}\cup \{0\} .
		\end{align*}
	\end{enumerate}
\end{proposition}
The proof for the above proposition is in Appendix \ref{app:pf_properties}.
It is noteworthy that the above proposition can be restated with $\mathcal{M}$ and $r$ replaced by $\mathcal{M}_k$ and $r_k$, respectively, which are given by
$$\mathcal{M}_{k}=\mathcal{S}(\mathcal{M}_{k-1},r_k),~r_k \in [0,1),~k=1,2,\dots$$
Based on these properties of conelike neighborhoods, it is straightforward to verify that for an FDLTI system $P\in\mathcal{P}$, it holds that
\begin{align}\label{eq:gap_in_conelike}
\bigcup\left\{\mathcal{G}^2_{\tilde{P}}:~ \tilde{P} \in \mathcal{B}(P,r)\right\} \subset \mathcal{S}(\mathcal{G}^2_P,r).
\end{align}
In other words, the gap ball of FDLTI systems is contained in the conelike neighborhood, given the same nominal system and radius. \textcolor{black}{Moreover, the conelike neighborhoods are also closely related to the nonlinear directed gap balls \cite{georgiou1997robustness}, which will be briefly introduced and utilized in the next subsection.}

In general, the equality ${\mathcal{S}}(\mathcal{M},r) = \tilde{\mathcal{S}}(\mathcal{M},r)$ does not hold with respect to an arbitrary closed set $\mathcal{M}\subset\mathcal{L}_2$. This emphasizes the importance of restricting that the ``center'' of the neighborhood is a closed subspace $\mathcal{M}_0$, which may be interpreted as the $\mathcal{L}_2$ graph of an FDLTI system in the following developments.

%
As inspired by the standard gap metric result on FDLTI systems in Lemma~\ref{lem:gaprobust}, we have the following result concerning closed-loop systems subject to nonlinear perturbations, whose proof is provided in Appendix~\ref{app:pf_propo2}.
\begin{proposition}\label{prop:extend_arcsin}
	Let $\lcls\in\mathcal{RH}_\infty$ and $r_p,r_c \in [0,1)$. Then the following statements are equivalent.
	\begin{enumerate}
		\item [(a)] $\bmF_{\tilde{\bm{P}},\tilde{\bm{C}}}$ has a \textcolor{black}{bounded} inverse on $\mathcal{R}(\bmF_{\tilde{\bm{P}},\tilde{\bm{C}}})$ for all $\tilde{\bm{P}}$ with $\mathcal{G}^2_{\tilde{\bm{P}}} \subset \mathcal{S}(\mathcal{G}^2_{{P}},r_p)$ and $\tilde{\bm{C}}$ with
		$\mathcal{G}'^2_{\tilde{\bm{C}}}\subset \mathcal{S}(\mathcal{G}'^2_{{C}},r_c)$;
		\item [(b)] $\mathcal{S}(\mathcal{G}^2_{{P}},r_p)\cap \mathcal{S}(\mathcal{G}'^2_{{C}},r_c) = \{0\};$
		\item [(c)] $\arcsin r_p +\arcsin r_c < \arcsin \|P\,\#\,C\|^{-1}_\infty.$
	\end{enumerate}
\end{proposition}
In the above proposition, we present a necessary and sufficient ``pre-stability'' condition in terms of an ``arcsine'' inequality, allowing simultaneous nonlinear perturbations on the plant and controller. Under the condition of statement~(a), as long as we can further show that $\mathcal{R}(\bmF_{\tilde{\bm{P}},\tilde{\bm{C}}})=\mathcal{L}_2$, then the closed-loop stability of $\tilde{\bm{P}}\,\#\,\tilde{\bm{C}}$ follows from Lemma~\ref{lem:finite_gain_stability_def}.

It is worth noting that
for nonlinear systems, $\delta$-type gaps and $\gamma$-type gaps can be used to characterize uncertain systems through their graphs \cite{georgiou1997robustness,james2005gap}. In contrast, a conelike neighborhood simply gathers all input-output
pairs lying within a prespecified angular distance from its center --- the $\mathcal{L}_2$ graph of the nominal system. It is preferable to concentrate on only input-output pairs rather than system graphs since in many cases, only partial information about the graph of a nonlinear system is available. The given information may not be sufficient for the purpose of computing the gap-distance, thus rendering standard gap-type stability conditions inapplicable. On
the contrary, if the uncertainties are measured with respect to the available input-output pairs, it is likely that the limited measurements are
sufficient to give a good approximation of these uncertainties. To verify whether a partially known perturbed system lies within a conelike
neighborhood, it suffices to check every available energy-bounded input-output pair.

\subsection{Proof of the Main Theorem}
Before proceeding to the proof of Theorem \ref{thm:NL_main}, we introduce several useful lemmas in the following.

A linear version of Theorem~\ref{thm:NL_main} was obtained in \cite{di2020tac}, where two-port networks are modelled by finite-dimensional FDLTI systems, i.e.,
$$T_k=I+\Delta_k~\text{and}~\Delta_k \in \mathcal{RH}_\infty,~k=1,2,\dots,l.$$
The result \cite[Theorem~1]{di2020tac} is stated in the following lemma.
\begin{lemma}\label{lem:LTI_two-port_stability}
	Let $\lcls\in\mathcal{RH}_\infty$ and $ r_p, r_c, r_k \in [0,1)$. The NCS in Fig.~\ref{fignetwork} is stable for all $\tilde{P} \in \mathcal{B}(P, r_p), ~\tilde{C} \in \mathcal{B}(C, r_c)$ and $\Delta_k \in \mathcal{RH}_\infty$ with $\|\Delta_k\|_\infty \leq r_k, ~k = 1,2,\dots,l$, if and only if
	\vspace{-5pt}
	\begin{multline*}
	\arcsin r_p + \arcsin r_c + \sum _{k=1}^{l} \arcsin r_k \\
	< \arcsin \|P\,\#\,C\|_\infty^{-1}.
	\end{multline*}
\end{lemma}
The above lemma provides us one direct way to show the necessity of the robust stability condition~\eqref{eq:NL_two-port_arcsin}, as will be elaborated momentarily. In order to show its sufficiency, we first introduce the following lemma that characterizes inclusive relations of conelike sets. The lemma employs similar expressions to the relations induced by the angular metric \cite{qiu1992feedback}.
\begin{lemma}\label{lem:cascaded}
	Let $\mathcal{M}_0\subset \mathcal{L}_2$ be a closed subspace and $\mathcal{M}_{j}=\mathcal{S}(\mathcal{M}_{j-1},r_j)$ for $r_j \in [0,1)$, $j=1,2,\dots,k$, satisfying $\sum_{j=1}^k \arcsin r_j\leq \pi/2$. Then it holds that
	$$\mathcal{M}_{k}\subset\mathcal{S}\left(\mathcal{M}_0,\sin\left(\sum_{j=1}^k \arcsin r_j\right)\right).$$
\end{lemma}
\begin{pf}
	It suffices to show that
	$$\mathcal{M}_2\subset\mathcal{S}(\mathcal{M}_0,\sin(\arcsin r_1 + \arcsin r_2)),$$
	and the rest follows by induction.
	
	For any $u \in \mathcal{M}_0$, $u_1 \in \mathcal{M}_1$ and $u_2 \in \mathcal{M}_2$, we have
	\begin{align}\label{ineq:angles2}
	\theta(u_2,u) \leq \theta(u_2,u_1) + \theta(u_1,u).
	\end{align}
	Particularly, let $\bar{u}_1 \in \argmin_{u_1 \in \mathcal{M}_1} \theta(u_2,u_1)$. Then inequality (\ref{ineq:angles2}) implies that
	\begin{align*}
	\theta(u_2,u) &\leq \theta(u_2,\bar{u}_1) + \theta(\bar{u}_1,u) \\
	&\leq \arcsin r_2 + \theta(\bar{u}_1,u).
	\end{align*}
	Taking infimums on the both sides and noting Proposition~\ref{prop:properties_conelike}(b), we obtain that
	\begin{align*}
	\inf_{u \in \mathcal{M}_0}\theta(u_2,u) &\leq \arcsin r_2 + \inf_{u \in \mathcal{M}_0}\theta(\bar{u}_1,u) \\
	&\leq \arcsin r_1 + \arcsin r_2.
	\end{align*}
	Hence, $u_2 \in \mathcal{S}(\mathcal{M}_0,\sin(\arcsin r_1 + \arcsin r_2))$ by another application of Proposition~\ref{prop:properties_conelike}(b).\hspace*{\fill}\qed
\end{pf}

In what follows, we introduce an important robust stability result related to the directed nonlinear gap \cite{georgiou1997robustness}. For systems $\bmP_1$ and $\bmP_2$, define the directed nonlinear gap from $\bmP_1$ to $\bmP_2$ as
\begin{multline}\label{eq:direct_gap}
\hspace{-3pt}\vec{\delta}(\bmP_1,\bmP_2)\hspace{-2pt}:=\hspace{-2pt}\lim\hspace{-1pt}\sup_{\hspace*{-15pt}T>0}\hspace{-2pt}\sup_{v\in\mathcal{G}_{\bm{P}_2}}\hspace{-2pt}\inf_{\substack{u\in\mathcal{G}_{\bm{P}_1},\\ \|\bm{\Gamma}_T u\|_2\neq 0}}\hspace{-6pt}\frac{\|\bmG_T(u\hspace{-1pt}-\hspace{-1pt}v)\|_2}{\|\bmG_T u\|_2}.
\end{multline}
The following lemma is adapted from \cite[Theorem~3]{georgiou1997robustness}.
\begin{lemma}\label{lem:NL_gap_stable}
	Let nonlinear system $\nlcls$ be finite-gain stable. Then $\tilde{\bmP}\,\#\,\bmC$ is finite-gain stable for all $\tilde{\bmP}$ with $$\vec{\delta}({\bmP},\tilde{\bmP})<\|\bm{\Pi}_{\mathcal{G}_{\bm{P}}\sslash\mathcal{G}'_{\bm{C}}}\|^{-1}.$$
\end{lemma}
Furthermore, we have the following important lemma relating cascaded two-port uncertainty neighborhoods to the directed nonlinear gap. Given an FDLTI closed-loop system $\lcls$ and incrementally stable uncertainty quartets $\bmD_j$, $j=1,2,\dots,l$, we define a family of perturbed plants and controllers $\tilde{\bm{P}}_k(\tau)$ and $\tilde{\bm{C}}_k(\nu)$ parameterized by $\tau$ and $\nu\in[0,1]$, respectively, via
\begin{align*}
&\mathcal{G}_{\tilde{\bm{P}}_k(\tau)} = (\bm{I}+\tau\bm{\Delta}_{k})\cdots (\bm{I}+\tau\bm{\Delta}_{1})\mathcal{G}_{P},\\
&\mathcal{G}'_{\tilde{\bm{C}}_k(\nu)} = (\bmI+\nu\bm{\Delta}_{k+1})^{-1}\cdots (\bmI+\nu\bm{\Delta}_{l})^{-1}\mathcal{G}'_{C}.
\end{align*}
\begin{lemma}\label{lem:homotopy}
	For all $\epsilon>0$, there exists $\delta>0$ such that for all $\tau,\nu\in[0,1)$, it holds that
	\begin{align*}
	&\vec{\delta}(\tilde{\bm{P}}_k(\tau),\tilde{\bm{P}}_k(\tau+\delta))<\epsilon~\text{and}~\vec{\delta}(\tilde{\bm{C}}_k(\nu),\tilde{\bm{C}}_k(\nu+\delta))<\epsilon.
	\end{align*}
\end{lemma}
In other words, $\tilde{\bm{P}}_k(\tau)$ and $\tilde{\bm{C}}_k(\nu)$, as functions of $\tau$ and $\nu$, respectively, are uniformly continuous with respect to the directed gap.
\begin{pf}
	We prove the lemma for $\tilde{\bmP}_k(\tau)$, and the proof for $\tilde{\bmC}_k(\nu)$ follows from similar arguments. Let $L$ be a common Lipschitz constant for all $\bmD_j$, $j=1,2,\dots,k$.
	Let $x_0\in\mathcal{G}_{P}$, and for $j=1,2,\dots,k$,
	\begin{align*}&x_j=(\bmI+\tau\bmD_j)\cdots(\bmI+\tau\bmD_1)x_0,\\&\tilde{x}_j=(\bmI+(\tau+\delta)\bmD_j)\cdots(\bmI+(\tau+\delta)\bmD_1)x_0.\end{align*}
	Hence it is clear that
	\begin{align}\label{eq:pf_lem_homotopy_0}
	\hspace{-8pt}\|\bmG_Tx_k\|_2\geq (1-r_1)\cdots(1-r_k)\|\bmG_T x_0\|_2,\forall~T\geq 0.
	\end{align}
	Furthermore, we claim that
	\begin{align}\label{eq:pf_lem_homotopy}
	\|\bmG_T(x_k-\tilde{x}_k)\|_2\leq \alpha\delta\|\bmG_Tx_0\|_2, ~\forall~T\geq 0,
	\end{align}
	where $\alpha>0$ is a constant that is independent of $\tau$, $\delta$ and $x_0$. To see this, observe that
	$$\|\bmG_T(x_1-\tilde{x}_1)\|_2=\|\bmG_T\delta\bmD_1x_0\|_2\leq r_1\delta\|\bmG_Tx_0\|_2.$$
	Moreover, by the Lipschitz continuity of $\bmD_j$, we have
	\begin{align*}
	&\|\bmG_T(x_2-\tilde{x}_2)\|_2\\
	&=\|\bmG_T((\bmI+\tau\bmD_2)x_1-(\bmI+\tau\bmD_2)\tilde{x}_1-\delta\bmD_2\tilde{x}_1)\|_2\\
	&\leq\|\bmG_T((\bmI+\tau\bmD_2)x_1-(\bmI+\tau\bmD_2)\tilde{x}_1)\|_2\hspace{-2pt}+\hspace{-2pt}\|\bmG_T(\delta\bmD_2\tilde{x}_1)\|_2\\
	&\leq (1+\tau L)\|\bmG_T(x_1-\tilde{x}_1)\|_2+\delta r_2\|\bmG_T\tilde{x}_1\|_2\\
	&\leq ((1+L)r_1+r_2(1+r_1))\delta \|\bmG_Tx_0\|_2.
	\end{align*}
	By repeating the arguments above iteratively, we arrive at the claim in \eqref{eq:pf_lem_homotopy} as required. Combining this with \eqref{eq:pf_lem_homotopy_0}, we obtain that for all $T>0$,
	\begin{align*}
	\sup_{v\in\mathcal{G}_{\tilde{\bm{P}}_k(\tau+\delta)}}&\inf_{\substack{u\in\mathcal{G}_{\tilde{\bm{P}}_k(\tau)},\\ \|\bm{\Gamma}_T u\|_2\neq 0}}\frac{\|\bmG_T(u-v)\|_2}{\|\bmG_T u\|_2}\\
	&\leq \sup_{\substack{x_0\in\mathcal{G}_{P},\\\|\bm{\Gamma}_Tx_o\|_2\neq 0}}\frac{\|\bmG_T(x_k-\tilde{x}_k)\|_2}{\|\bmG_T x_k\|_2}\\
	&\leq \sup_{\substack{x_0\in\mathcal{G}_{P},\\\|\bm{\Gamma}_Tx_o\|_2\neq 0}} \frac{\alpha\delta\|\bmG_Tx_0\|_2}{(1-r_1)\cdots(1-r_k)\|\bmG_Tx_0\|_2}\\
	&=\frac{\alpha\delta}{(1-r_1)\cdots(1-r_k)}.
	\end{align*}
	Setting
	$$\delta=\epsilon(1-r_1)\cdots(1-r_k)/\alpha>0,$$
	and noting the definition of the directed nonlinear gap in \eqref{eq:direct_gap}, we complete the proof.\hspace*{\fill}\qed
\end{pf}

Based on the above lemmas, we are ready to prove Theorem \ref{thm:NL_main} as follows. \textcolor{black}{The proof borrows the idea of using gap-metric homotopy to establish feedback stability from \cite{rantzer1997integral,cantoni2012robustness}.}

\begin{pf}
	First we show the necessity using contrapositive arguments. Assume that inequality~(\ref{eq:NL_two-port_arcsin}) does not hold.
	Then it follows from Lemma~\ref{lem:LTI_two-port_stability} that there exist systems $\tilde{P}$, $\tilde{C}\in\mathcal{P}$ and stable uncertainties $\Delta_k\in\mathcal{RH}_\infty$, $k=1,2,\dots,l$ satisfying that $\tilde{P} \in \mathcal{B}(P, r_p), ~\tilde{C} \in \mathcal{B}(C, r_c)$, and $\|\Delta_k\|_\infty \leq r_k$, so that
	$${P}_q\,\#\,{C}_q \notin\mathcal{H}_\infty$$
	for an integer $q\in [0,l]$. Here, $P_q, C_q$ are determined, respectively, by
	\begin{equation*}
	\begin{aligned}
	\mathcal{G}_{{{P}}_q} &= ({I}+{{\Delta}}_{q})\cdots ({I}+{{\Delta}}_{1})\mathcal{G}_{\tilde{P}},\\
	\mathcal{G}'_{{{C}}_q} &= (I+{{\Delta}}_{q+1})^{-1}\cdots (I+{{\Delta}}_{l})^{-1}\mathcal{G}'_{\tilde{C}}.
	\end{aligned}
	\end{equation*}
	\textcolor{black}{Since every strongly causal FDLTI system admits a strictly proper transfer function representation, it follows from the hypothesis that either $P$ or $C$ is strictly proper. Therefore, one can verify that there exists an $\hat{\omega}\neq \infty$ such that $\hat{\omega}\in\argmax_{\omega\in\mathbb{R}\cup\infty}\bar{\sigma}(P(j\omega)\,\#\,C(j\omega))$. By the proof of necessity of Lemma~\ref{lem:LTI_two-port_stability} from \cite{di2020tac} and using the interpolation method in \cite[Lemma~1.14]{vinnicombe2000uncertainty}\footnote{Notably, \cite[Lemma~1.14]{vinnicombe2000uncertainty} solves the interpolation problem for the case when $\hat{\omega}\in(0,\infty)$. When $\hat{\omega}=0$, it suffices to multiply $1/(1+s)$ onto $\Delta$ to obtain the desired strictly proper real-rational interpolation.} for $\hat{\omega}$, we can further require that $\Delta_k\in\mathcal{RH}_\infty$, $k=1,2,\dots,l$, are strictly proper transfer matrices, i.e., they are strongly causal.}
	Therefore, by contraposition and Proposition~\ref{prop:stability}, we prove the necessity of the robust stability condition in Theorem~\ref{thm:NL_main}.
	
	In the rest of this proof, we show the sufficiency of the robust stability condition in three steps.\\
	\textit{Step~1}: Suppose that we are at the $k$-th stage of equivalent closed-loop systems as shown in Fig.~\ref{fig:equiv_con_pla}. Let
	$$\mathcal{M} = \mathcal{G}^2_{{P}},~\tilde{\mathcal{M}}_0  = \mathcal{G}^2_{\tilde{P}},~\tilde{\mathcal{M}}_j(\tau) = \mathcal{G}^2_{\tilde{\bm{P}}_j(\tau)},~j=1,2,\dots,k,$$ where $\mathcal{G}_{\tilde{\bm{P}}_j(\tau)} = (\bm{I}+\tau\bm{\Delta}_{j})\cdots (\bm{I}+\tau\bm{\Delta}_{1})\mathcal{G}_{\tilde{P}}.$
	Then it follows that
	\begin{align*}
	&\mathcal{\tilde{M}}_0\subset \mathcal{S}(\mathcal{M},r_p),\\
	&\tilde{\mathcal{M}}_{j}(\tau) = (\bm{I}+\tau\bm{\Delta}_j)\tilde{\mathcal{M}}_{j-1}(\tau),
	\end{align*}
	with $\|\tau\bm{\Delta}_j\|\leq\tau r_j$. Let $ v \in \tilde{\mathcal{M}}_{j}(\tau) \setminus \{0\}$, then there exists $u_1 \in \tilde{\mathcal{M}}_{j-1}(\tau)$ such that $v=(\bm{I}+\tau\bm{\Delta}_j)u_1$. Hence we have
	$$\inf_{0\neq u \in \tilde{\mathcal{M}}_{j-1}(\tau)}\frac{\|v-u\|_2}{\|u\|_2} \leq \frac{\|\tau\bm{\Delta}_j u_1\|_2}{\|u_1\|_2} \leq \|\tau\bm{\Delta}_j\|\leq\tau r_j.$$
	As a result,
	$\tilde{\mathcal{M}}_j(\tau) \subset \mathcal{S}(\tilde{\mathcal{M}}_{j-1}(\tau),\tau r_j)$, $j=1,2,\dots,k$.
	Since $\tilde{\mathcal{M}}_0$ is a closed subspace in $\mathcal{L}_2$, it follows from Lemma \ref{lem:cascaded} and \eqref{eq:gap_in_conelike} that
	\begin{align*}
	\tilde{\mathcal{M}}_k(\tau) &\subset \mathcal{S}\left(\tilde{\mathcal{M}}_0,\sin\left(\sum_{j=1}^{k} \arcsin \tau r_j\right)\right)\\ &\subset \mathcal{S}\left({\mathcal{M}},\sin\left(\arcsin r_p + \sum_{j=1}^{k} \arcsin \tau r_j\right)\right).
	\end{align*}
	Likewise, for the controller part, let
	$$\mathcal{N} = \mathcal{G}'^2_{{C}},~\tilde{\mathcal{N}}_l = \mathcal{G}'^2_{\tilde{{C}}},~\tilde{\mathcal{N}}_j(\nu) = \mathcal{G}'^2_{\tilde{\bm{C}}_j(\nu)},j=k+1,\dots,l,$$
	where
	$\mathcal{G}'_{\tilde{\bm{C}}_j(\nu)} = (\bmI+\nu\bm{\Delta}_{j+1})^{-1}\cdots (\bmI+\nu\bm{\Delta}_{l})^{-1}\mathcal{G}'_{\tilde{C}}.$
	Then it follows that
	\begin{align*}
	&\mathcal{\tilde{N}}_l\subset \mathcal{S}(\mathcal{N},r_c),\\
	&\tilde{\mathcal{N}}_{j-1}(\nu) = (\bm{I}+\nu\bm{\Delta}_j)^{-1}\tilde{\mathcal{N}}_{j}(\nu),
	\end{align*}
	with $\|\nu\bm{\Delta}_j\|\leq\nu r_j$. Let $ v \in \tilde{\mathcal{N}}_{j-1}(\nu)\setminus \{0\}$, then there exists $u_1 \in \tilde{\mathcal{N}}_{j}(\nu)$ such that $v=(\bm{I}+\nu\bm{\Delta}_j)^{-1}u_1$. Hence we have
	$$\inf_{0\neq u \in \tilde{\mathcal{N}}_{j}(\nu)}\frac{\|v-u\|_2}{\|v\|_2} \leq \frac{\|\nu\bm{\Delta}_j v\|_2}{\|v\|_2} \leq \|\nu\bm{\Delta}_j\|\leq \nu r_j.$$
	Thus for $j = k+1,k+2,\dots,l$, we have that
	$$\tilde{\mathcal{N}}_{j-1}(\nu) \subset \tilde{\mathcal{S}}(\tilde{\mathcal{N}}_{j}(\nu),\nu r_j) = {\mathcal{S}}(\tilde{\mathcal{N}}_{j}(\nu),\nu r_j),$$
	where the equality follows from Proposition~\ref{prop:properties_conelike}(b).
	Applying Lemma~\ref{lem:cascaded} and \eqref{eq:gap_in_conelike} yields that
	$$\tilde{\mathcal{N}}_k(\nu) \subset \mathcal{S}\left(\mathcal{N},\sin\left( \arcsin r_c + \sum_{j=k+1}^{l} \arcsin\nu r_j\right)\right).$$
	Therefore, from the stability condition (\ref{eq:NL_two-port_arcsin}) and Proposition~\ref{prop:extend_arcsin}, we know $\bmF_{\tilde{\bm{P}}_k(\tau),\tilde{\bm{C}}_k(\nu)}$ has a uniformly bounded inverse on $\mathcal{R}(\bm{F}_{\tilde{\bm{P}}_k(\tau),\tilde{\bm{C}}_k(\nu)})$ over all $\tau,\nu\in[0,1]$, which ensures the existence of a constant $\alpha>0$ such that
	\begin{align}\label{eq:pf_parallel_bounded}
	\left\|\bm{\Pi}_{\mathcal{G}_{\tilde{\bm{P}}_j(\tau)}\sslash{\mathcal{G}'_{\tilde{\bm{C}_j}(\nu)}}}x\right\|_2\leq \alpha\left\|x\right\|_2,
	\end{align}
	for all $x\in\mathcal{R}(\bm{F}_{\tilde{\bm{P}}_k(\tau),\tilde{\bm{C}}_k(\nu)})$ and $\tau,\nu\in[0,1]$.
	
	\textit{Step 2:} Note that $\tilde{\bm{P}}_k(\tau)$ and  $\tilde{\bm{C}}_k(\nu)$ are uniformly continuous with respect to the directed nonlinear gap. In particular, it follows from Lemma~\ref{lem:homotopy} that for $\alpha>0$ given in \eqref{eq:pf_parallel_bounded}, there exists $\delta>0$ such that
	\begin{multline}\label{eq:pf_homotopy_step}
	\vec{\delta}(\tilde{\bm{P}}_k(\tau),\tilde{\bm{P}}_k(\tau+\delta))<\frac{1}{\alpha},\\
	\text{and}~\vec{\delta}(\tilde{\bm{C}}_k(\nu),\tilde{\bm{C}}_k(\nu+\delta))<\frac{1}{\alpha},
	\end{multline}
	for all $\tau,\nu\in[0,1)$.
	
	\textit{Step 3:} When $\tau=\nu=0$, it follows from Lemma~\ref{lem:gaprobust} that $$\tilde{\bm{P}}_k(0)\,\#\,\tilde{\bm{C}}_k(0)=\tilde{P}\,\#\,\tilde{C}$$ is stable. Hence \eqref{eq:pf_parallel_bounded} implies that $$\left\|\bm{\Pi}_{\mathcal{G}_{\tilde{\bm{P}}_j(0)}\sslash{\mathcal{G}'_{\tilde{\bm{C}_j}(0)}}}\right\|\leq \alpha.$$
	Then combining \eqref{eq:pf_homotopy_step} with Lemma~\ref{lem:NL_gap_stable}, we obtain the finite-gain stability of $\tilde{\bm{P}}_k(\delta)\,\#\,\tilde{\bm{C}}_k(0)$. By iteratively using \eqref{eq:pf_parallel_bounded}, \eqref{eq:pf_homotopy_step} and Lemma~\ref{lem:NL_gap_stable}, we obtain that all the closed-loop systems in the following sequence are finite-gain stable:
	\begin{multline*}
	\tilde{\bm{P}}_k(\delta)\,\#\,\tilde{\bm{C}}_k(\delta),~\tilde{\bm{P}}_k(2\delta)\,\#\,\tilde{\bm{C}}_k(\delta),~\tilde{\bm{P}}_k(2\delta)\,\#\,\tilde{\bm{C}}_k(2\delta),\\
	\dots,\tilde{\bm{P}}_k(1)\,\#\,\tilde{\bm{C}}_k(1-\delta),~\tilde{\bm{P}}_k(1)\,\#\,\tilde{\bm{C}}_k(1).
	\end{multline*}
	The finite-gain stability of the two-port NCS then follows by Proposition~\ref{prop:stability}.\hspace*{\fill}\qed
\end{pf}

\section{Conclusions and Future Work}\label{sec:conc}
We investigate the robust stabilization problem of a two-port NCS where the plant and controller are subject to gap-type uncertainties and the two-port communication channels are subject to nonlinear perturbations.
In order to characterize nonlinear uncertainty, a special conelike neighborhood, which is inspired and motivated by the elegant geometric properties of the gap metric, is introduced and investigated. A necessary and sufficient robust stability condition for the two-port NCS is given in the form of an ``arcsine'' inequality. The associated robust controller synthesis problem can be settled by solving an $\mathcal{H}_\infty$ control problem.

The development of the main result in this paper is based on a geometric approach, where the angles between conelike neighborhoods play a crucial role. This approach may be further applied to other related problems involving closed-loop stability. One can generalize the current problem setup by modeling communication channels as two-port networks with various types of interconnections, such as cascade connections, parallel connections, series connections, hybrid connections and so on. \textcolor{black}{In terms of technical developments, we can extend the current model of two-port NCSs in the language of the behaviour approach \cite{willems1998behaviour} so as to overcome the difficulty in modelling non-invertible equipments, such as quantizers. }


\bibliographystyle{apacite}    
\bibliography{TwoPortarX}           

\appendix
\section{Proof of Proposition~\ref{prop:stability}}\label{app:pf_stability}
\begin{pf}
	The necessity is obvious. We show sufficiency below. Define a \textcolor{black}{sequence} of operators $\bm{B}_k= I_k \mapsto [v_k~w_k]^T$ and $\tilde{\bm{B}}_k= I_k \mapsto [u_k~y_k]^T$, $k=0,1,\dots,l$.
	Let $\tilde{\bm{P}}_k\,\#\,\tilde{\bm{C}}_k$ be finite-gain stable, and thus $\bm{B}_k$ and $\tilde{\bm{B}}_k$ are finite-gain stable. As $\|\bm{\Delta}_k\| < 1$ and the instantaneous gain of $\bmD_k$ is less than 1, $k=1,2,\dots,l$, the finite-gain stability of $(\bm{I}+\bm{\Delta}_k)^{-1}$ follows by the nonlinear small-gain theorem. Hence the composite operators
	\begin{align*}&(\bm{I}+\bm{\Delta}_{k+1})\bm{B}_k=I_k \mapsto \begin{bmatrix}v_{k+1}~~w_{k+1}\end{bmatrix}^T,\\
	&(\bm{I}+\bm{\Delta}_k)^{-1}\tilde{\bm{B}}_k=I_k \mapsto \begin{bmatrix}u_{k-1}& y_{k-1}\end{bmatrix}^T\end{align*}
	are finite-gain stable. By iterative compositions of the maps, we obtain the finite-gain stability of $\bm{A}_k$ for all $k = 0, 1, \ldots l$.\hspace*{\fill}\qed
\end{pf}
\section{Proof of Proposition~\ref{prop:properties_conelike}}\label{app:pf_properties}
\begin{pf}
From the closedness of the conelike neighborhood $\mathcal{M} \subset \mathcal{L}_2$, we can replace ``infimum'' with ``minimum'' in the definition of $\tilde{\mathcal{S}}(\mathcal{M},r)$ and ${\mathcal{S}}(\mathcal{M},r)$.
	
For statement (a), we first notice that $0\in\mathcal{M}$ and $u \in \mathcal{M} \Rightarrow \alpha u \in \mathcal{M}$ with $\alpha\neq 0$, which follows from that $\mathcal{M}_0$ is a closed subspace and that
		$$\min_{0\neq x \in \mathcal{M}_0}\frac{\|\alpha u-x\|}{\|x\|} = \min_{0\neq \alpha x \in \mathcal{M}_0}\frac{\|\alpha u-\alpha x\|}{\|\alpha x\|} \leq r_0. $$
		Next we show $v \in \mathcal{S}(\mathcal{M},r)\Rightarrow\alpha v \in \mathcal{S}(\mathcal{M},r)$ with $\alpha \in \mathbb{R}$.
		Recall the definition of the neighborhood, then we have
	$$\min_{0\neq u \in \mathcal{M}}\frac{\|\alpha v-u\|}{\|u\|} = \min_{0\neq \alpha u \in \mathcal{M}}\frac{\|\alpha v-\alpha u\|}{\|\alpha u\|} \leq r,$$
	which proves statement (a).
	
	For statement (b), first we show that
	$$\mathcal{S}(\mathcal{M},r)
	=\left\{v\in \mathcal{L}_2 :~\min_{u\in \mathcal{M}} \theta(u,v)\leq \arcsin r\right\}\cup \{0\} .$$
	Let $v \in \mathcal{S}(\mathcal{M},r)$ and
	$$\bar{u} \in \argmin_{u\in \mathcal{M}} \frac{\|v-u\|_2}{\|u\|_2}.$$
	\textcolor{black}{Then $v-\bar{u} \perp v$, which implies $$\sin \theta(\bar{u},v) = \dfrac{\|v-\bar{u}\|_2}{\|\bar{u}\|_2} \leq r.$$
		Consequently,
		$$\mathcal{S}(\mathcal{M},r) \subset \left\{v\in \mathcal{L}_2 : \min_{u\in \mathcal{M}} \theta(u,v)\leq \arcsin r\right\}\cup \{0\}.$$
		 On the other hand, let $v$ belong to the latter set. From statement (a), we can find with constant appropriate scaling a $\bar{u} \in \mathcal{S}(\mathcal{M},r)$ such that $\theta(\bar{u},v) \leq \arcsin r $ and $v-\bar{u} \perp v$, which implies that }
	$$\frac{\|v-\bar{u}\|_2}{\|\bar{u}\|_2} = \sin \theta(\bar{u},v)\leq r.$$
	It follows that $v\in\mathcal{S}(\mathcal{M},r)$.
	
\begin{figure}
		\centering
		\includegraphics[width=.36\textwidth]{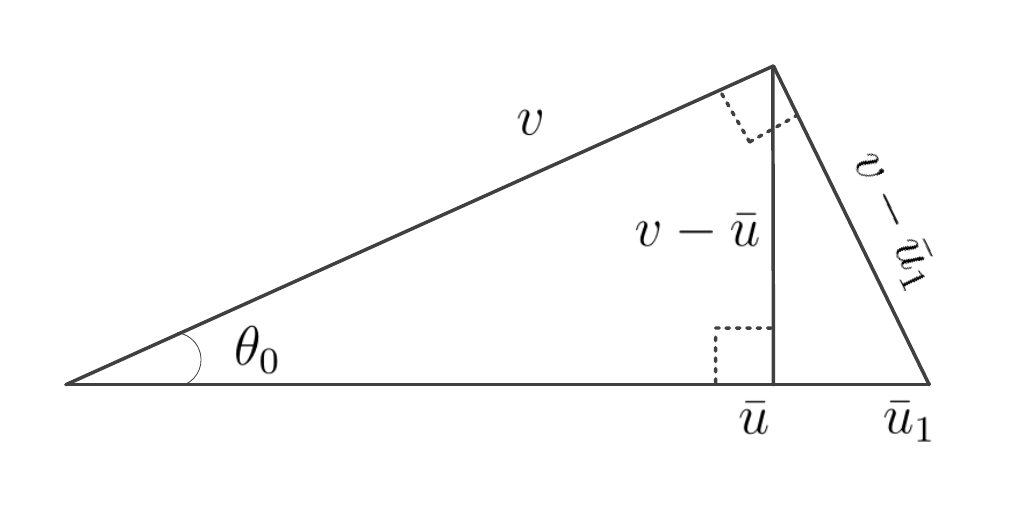}\\
		\caption{An illustrative diagram for the proof of Proposition~\ref{prop:properties_conelike}.}\label{fig:conelike}
	\end{figure}
Next we show that $\mathcal{S}(\mathcal{M},r) = \tilde{\mathcal{S}}(\mathcal{M},r).$
Let $v \in \tilde{\mathcal{S}}(\mathcal{M},r)$ and
	$$\bar{u} \in \argmin_{ u\in \mathcal{M}} \frac{\|v-u\|_2}{\|v\|_2}.$$
Then $v-\bar{u} \perp \bar{u}$. Denote the acute angle between $v$ and $\bar{u}$ as $\theta_0$ $(\leq \arcsin r)$. In the hyperplane determined by $\bar{u}$ and $v$, as shown in Fig.~\ref{fig:conelike}, we can scale $\bar{u}$ to $\bar{u}_1 \in \mathcal{M}$ along $\bar{u}$ so that $v-\bar{u}_1 \perp v$, which is allowable by statement (a). This implies that
$$\min_{0 \neq u\in \mathcal{M}} \frac{\|v-u\|_2}{\|u\|_2} \leq \frac{\|v-\bar{u}_1\|_2}{\|\bar{u}_1\|_2} = \sin \theta_0 \leq r.$$
Consequently, $\tilde{\mathcal{S}}(\mathcal{M},r)\subset {\mathcal{S}}(\mathcal{M},r)$.

That ${\mathcal{S}}(\mathcal{M},r)\subset \tilde{\mathcal{S}}(\mathcal{M},r)$ can be shown using similar arguments. This completes the proof.\hspace*{\fill}\qed
\end{pf}
\section{Proof of Proposition~\ref{prop:extend_arcsin}}\label{app:pf_propo2}
\begin{pf}
First we show that (c) implies (b).
Note from \cite{qiu1992feedback} that
\begin{align}\label{eq:pf_prop3}
\|P\,\#\,C\|^{-1}_\infty = \inf_{u\in \mathcal{G}^2_{{P}},v\in \mathcal{G}'^2_{{C}}} \sin \theta(u,v).
\end{align}
Let $u_1 \in \mathcal{S}(\mathcal{G}^2_{{P}},r_p)$ and $v_1 \in \mathcal{S}(\mathcal{G}'^2_{{C}},r_c)$.
It follows from Proposition~\ref{prop:properties_conelike} that
    \begin{align*}
    \inf_{u \in \mathcal{G}^2_{{P}}} \theta (u,u_1) \leq \arcsin r_p,~~\text{and}~~ \inf_{v \in \mathcal{G}'^2_{{C}}} \theta(v,v_1) \leq \arcsin r_c.
    \end{align*}
Since $\mathcal{G}^2_{P}$ and $\mathcal{G}'^2_{C}$ are closed subspaces, we let
    \begin{align*}\bar{u}\in \argmin_{u \in \mathcal{G}^2_{{P}}} \theta(u,u_1)~~\text{and}~~
	\bar{v}\in \argmin_{v \in \mathcal{G}'^2_{{C}}} \theta(v,v_1).\end{align*}
Then by the triangle inequality, we obtain that
	\begin{align*}
	\theta(u_1,v_1) &\geq \theta(\bar{u},\bar{v})-\theta(\bar{u},u_1)-\theta(\bar{v},v_1)\\
	&\geq \theta(\bar{u},\bar{v})-\arcsin r_p -\arcsin r_c\\
	&\geq \arcsin \|P\,\#\,C\|^{-1}_\infty-\arcsin r_p -\arcsin r_c\\
	&=: \epsilon > 0.
	\end{align*}
Hence it holds that
	$$\inf_{\substack{u_1 \in \mathcal{S}(\mathcal{G}^2_{{P}},r_p),\\v_1 \in \mathcal{S}(\mathcal{G}'^2_{{C}},r_c)}} \arcsin \theta (u_1,v_1) \geq \epsilon > 0,$$
	which implies $$\mathcal{S}(\mathcal{G}^2_{{P}},r_p)\cap \mathcal{S}(\mathcal{G}'^2_{{C}},r_c) = \{0\},$$
	as required.

Next we show that (b) implies (a) using contrapositive arguments. Let $\mathcal{G}^2_{\tilde{\bm{P}}}\subset \mathcal{S}(\mathcal{G}^2_{{P}},r_p),~\mathcal{G}'^2_{\tilde{\bm{C}}}\subset \mathcal{S}(\mathcal{G}'^2_{{C}},r_c)$,
	$\tilde{\bm{P}}\,\#\,\tilde{\bm{C}}$ be well-posed, but $\bm{\Pi}_{\mathcal{G}_{\tilde{\bm{P}}}\sslash\mathcal{G}'_{\tilde{\bm{C}}}}$ be unbounded on $\mathcal{R}(\bmF_{\tilde{\bm{P}},\tilde{\bm{C}}})$, i.e., there exists a sequence
	$\{\omega_k\}_{k=1}^{\infty} \subset \mathcal{R}(\bmF_{\tilde{\bm{P}},\tilde{\bm{C}}})\setminus \{0\}$ such that
	\begin{itemize}
		\item $\|\omega_k\|_2$, $k=1,2,\dots$, is increasing,
		\item $\displaystyle \lim _{k \to \infty}\|\omega_k\|_2=\infty~\text{and}~\lim _{k \to \infty}\frac{\|\bm{\Pi}_{\mathcal{G}_{\tilde{\bm{P}}}\sslash\mathcal{G}'_{\tilde{\bm{C}}}} \omega_k\|_2}{\|\omega_k\|_2} = \infty$.
	\end{itemize}
	From the well-posedness assumption, we know that
	$$\omega_k = \bm{\Pi}_{\mathcal{G}_{\tilde{\bm{P}}}\sslash\mathcal{G}'_{\tilde{\bm{C}}}}\omega_k + \bm{\Pi}_{\mathcal{G}'_{\tilde{\bm{C}}}\sslash\mathcal{G}_{\tilde{\bm{P}}}}\omega_k=: u_k+v_k.$$
    Since $\omega_k\in\mathcal{R}(\bmF_{\tilde{\bm{P}},\tilde{\bm{C}}})\subset\mathcal{L}_2$, it follows from Definition~\ref{def:wellposed} that $u_k,v_k\in\mathcal{L}_2\setminus\{0\}$ and
	$$\alpha_k := \dfrac{\|\omega_k\|_2}{\|u_k\|_2} \to 0$$ as $k \to \infty$.
	Consequently, we obtain that
	\begin{align*}
	\cos \theta(u_k,v_k)& = \left|\frac{\langle u_k,v_k\rangle}{\|u_k\|_2\|v_k\|_2}\right| \\
	& \geq \left|\frac{\langle u_k,\omega_k-u_k\rangle}{\|u_k\|_2(\|u_k\|_2+\|\omega_k\|_2)}\right| \\
	& \geq \frac{1}{1+\alpha_k}\left(1 - \frac{|\langle u_k,\omega_k\rangle|}{\|u_k\|^2_2}\right)\\
	& \geq \frac{1}{1+\alpha_k}\left(1-\frac{\|u_k\|_2\|\omega_k\|_2}{\|u_k\|^2_2}\right)\\
	& = \frac{1-\alpha_k}{1+\alpha_k} \to 1 ~~~~\text{as} ~k \to \infty.
	\end{align*}
	Note that $\theta(u_k,v_k)$ is an acute angle, and it holds that $\theta(u_k,v_k) \to 0$. Since $\mathcal{G}^2_{\tilde{\bm{P}}}$ and $\mathcal{G}'^2_{\tilde{\bm{C}}}$ are closed, it follows that
	$\mathcal{G}^2_{\tilde{\bm{P}}} \cap \mathcal{G}'^2_{\tilde{\bm{C}}}\neq \{0\}$, whereby
	$\mathcal{S}(\mathcal{G}^2_{P},r_p)\cap \mathcal{S}(\mathcal{G}'^2_{C},r_c) \neq \{0\}$.
	
	Finally, we prove that (a) implies (c) by contradiction.
	Suppose statement~(c) does not hold, i.e.,
	\[
	\arcsin r_p +\arcsin r_c \geq \arcsin \|P\,\#\,C\|^{-1}_\infty.
	\]
	Then it follows from Lemma~\ref{lem:gaprobust} and \eqref{eq:pf_prop3} that there exists  $u\in\mathcal{L}_2\setminus\{0\}$ such that
	$$\alpha u\in\mathcal{G}^2_{\tilde{P}}\cap\mathcal{G}'^2_{\tilde{C}},~\forall \alpha \in\mathbb{R}$$
	for some $\tilde{P}\in\mathcal{B}(P,r_p)$ and $\tilde{C}\in\mathcal{B}(C,r_c)$.
	Noting \eqref{eq:gap_in_conelike}, we obtain that
	$$\mathcal{G}^2_{\tilde{{P}}} \subset \mathcal{S}(\mathcal{G}^2_{{P}},r_p)~~\text{and}~~\mathcal{G}'^2_{\tilde{{C}}}\subset \mathcal{S}(\mathcal{G}'^2_{{C}},r_c).$$
	Clearly it holds that $0\in\mathcal{R}(\bmF_{\tilde{P},\tilde{C}})$. Since $0=u+(-u)$ with $u\in\mathcal{G}^2_{\tilde{P}}$ and $-u\in\mathcal{G}'^2_{\tilde{C}}$, there exist $v$ and $w\in\mathcal{L}_2$ such that
$$u=\begin{bmatrix}v \\ \tilde{P}v\end{bmatrix}=-\begin{bmatrix}\tilde{C}w \\ w\end{bmatrix}.$$
Note $u\in\mathcal{L}_2\setminus\{0\}$, and let $z=[v ~ w]^T\in\mathcal{L}_2\setminus\{0\}$. It then follows by Definition~\ref{def:wellposed} that
	$$\bmF_{\tilde{P},\tilde{C}}z=0.$$
	Since $\bmF_{\tilde{P},\tilde{C}}0=0$ as well, $\bmF_{\tilde{P},\tilde{C}}$ is not invertible and the well-posedness assumption is thus violated, which leads to a contradiction.
\hspace*{\fill}\qed
\end{pf}

%

\end{document}